\documentclass[a4paper,fleqn,usenatbib,useAMS]{mnras}

\usepackage{graphicx}	
\usepackage{amsmath}	
\usepackage{amssymb}	
\usepackage{multicol}        
\usepackage{bm}		
\usepackage{pdflscape}	
\usepackage{journals}
\usepackage{epstopdf}

\def\HI{H~{\sc i}\, }

\usepackage[T1]{fontenc}
\usepackage{ae,aecompl}

\usepackage{newtxtext,newtxmath}

\title[What made discy galaxies  giant?]{What made discy galaxies  giant?
}
\author[A. Saburova]{A. S. Saburova$^{1}$\thanks{E-mail:
saburovaann@gmail.com}
\\
$^1$ Sternberg Astronomical Institute, Moscow M.V. Lomonosov State University, Universitetskij pr., 13,  Moscow, 119991, Russia\\
}

\begin{document}
\label{firstpage}
\pagerange{\pageref{firstpage}--\pageref{lastpage}} \pubyear{2016}
\maketitle

\begin{abstract}
I studied giant discy galaxies with optical radii more than 30 kpc. The comparison of these systems with discy galaxies of moderate sizes revealed that they tend to have higher rotation velocities, B-band luminosities, \HI masses and dark-to-luminous mass ratios. The giant discs follow the trend $\log(M_{\rm HI})(R_{25})$ found for normal size galaxies. It indicates the absence of the peculiarities of evolution of star formation in these galaxies. The \HI mass to luminosity ratio of giant galaxies appears not to differ from that of normal size galaxies, giving evidences in favor of similar star formation efficiency. I also found that the bars and rings occur more frequently among giant discs. I performed mass-modelling of the subsample of 18 giant galaxies with available rotation curves and surface photometry data and constructed $\chi^2$ maps for the parameters of their dark matter haloes. These estimates indicate that giant discs tend to be formed in larger more massive and rarified dark haloes in comparison to moderate size galaxies. However giant galaxies do not deviate significantly from the relations between the optical sizes and dark halo parameters for moderate size galaxies. These findings  can rule out the catastrophic scenario of the formation of at least most of giant discs, since they follow the same relations as normal discy galaxies. The giant sizes of the discs can be due to the high radial scale of the dark matter haloes in which they were formed. 
\end{abstract}

\section{Introduction}\label{intro}
One of the most important problems in the theories of formation and evolution of galaxies is what physical processes and what characteristics have most significant effect on the observed properties. Are these factors internal (the total mass or the properties of the host dark halo) or external (environment)? The possible way to find this out is to pay attention to the unusual objects like giant discy galaxies that may deviate significantly by the properties that are clue in evolution of galaxies.

 Giant discy galaxies and their extreme representatives giant low surface brightness (LSB) galaxies are being known for 30 years (\citealt{Bothun1990}; \citealt{Bothun1987}). They have high total masses of $\sim 10^{12} M\sun$ and  discs of enormous radii of up to 150 kpc. The  formation of such systems represents a certain problem for hierarchical clustering concept in which the dark haloes hosting discy galaxies do not undergo major mergers.

I decided to look into the galaxies with more moderate, but still high sizes with higher surface brightnesses in comparison to giant LSBs (the selection criteria and the sample are described below). I had three reasons to choose them.  Firstly, due to the higher surface brightnesses these systems have more available data on the rotational amplitude in comparison to giant LSBs, for some of them even extended stellar kinematics is measured. These data are very important if one needs to determine the parameters of dark matter haloes (which is one of the aims of my study). Secondly, these galaxies are more numerous in comparison to giant LSBs and I can study the statistics of their characteristics properly.  Thirdly, it is essential to understand how the properties of giant high surface brightness (HSB) discy galaxies correspond to that of giant LSBs and normal moderate size discs. Could they be intermediate link between giant LSBs and 'normal' discy galaxies?  

Several efforts were already made in order to understand the formation scenario for giant discs (see e.g. \citealt{Courtois2015}; \citealt{Ogleetal2016}; \citealt{Reshetnikov2010}; \citealt{Kasparova2014}; \citealt{Lelli2010} and references therein). \cite{Courtois2015} studied a sample of the most \HI  massive and fastest rotating disc galaxies in the local universe and found possible connection of their unusual properties with the environment. According to their research the most massive \HI detected galaxies are located preferentially in filaments. \cite{Ogleetal2016} considered superluminous spiral galaxies and proposed that they could be a result of major merger of two spiral gas-rich  galaxies, or alternatively they   could be formed gradually  by accretion of cold gas. In this case the halo should have a relatively low mass to avoid accretion shocks which can influence against the settling of the gas in the outer parts of the discs. \cite{Ghosh2008} studied giant ring galaxy UGC7069 with a diameter of $\sim 115$ kpc and concluded that it could be produced by flyby interactions between galaxies. They considered the presence of such giant ring as indirect
proof that ring galaxies might evolve into giant low surface brightness galaxies like Malin1. 
The major merger scenario of the formation of giant discy galaxies was also discussed in \cite{Reshetnikov2010} and \cite{Lelli2010} (see also the references therein). By contrast \cite{Kasparova2014} studied giant LSB galaxy Malin2 and proposed that there is no need to involve the catastrophic scenario and its unusual structure could be due to the peculiar properties of its dark matter halo - high radial scale of the halo and probably poor gas environment at the time when the disc was formed. 

The aim of current paper is to study the statistics of the giant HSB galaxies, to obtain their dark matter halo parameters and to compare them to the galaxies with 'normal' disc sizes and if possible -- with giant LSBs. It will aid in understanding of the reasons of the formation of  giant discy systems. 

The current paper is organized as follows: Section \ref{Obs} is devoted to the description of the selection criteria and the sample of giant spiral galaxies; the statistics of the parameters of giant galaxies and their comparison with the total sample of spiral galaxies from Hyperleda are given in Section \ref{Stat}; Section \ref{Parameters} is dedicated to the details of the estimate of the parameters of dark matter haloes of several giant galaxies;  the discussion is given in Section \ref{Discussion}; the main results are summarized in Section \ref{conclusion}.

\section{The sample }\label{Obs}
I consider the galaxies with the highest optical radii available in Hyperleda database\footnote{http://leda.univ-lyon1.fr/}, \citep{Makarovetal2014}.
 To define the sample of giant discy galaxies I requested the galaxies with half of the length of the projected major axis at the isophotal level 25 mag$/arcsec^2$ in the B-band $R_{25}$>30 kpc; with morphological types $t$>0; inclination $i$>40\degr with available measurements of rotation velocity. I added the latter two specifications since I was interested in the dynamical mass estimates of the galaxies. The low inclination angle makes the rotation velocity estimate uncertain, thus the galaxies which were close to face-on were removed by the selection criterion. Another source of the uncertainty of dynamical mass is in the non-circular velocities induced by the gravitational interaction of galaxies. The tight interaction can also lead to the erroneous determination of the optical radius. Thus, I inspected the images of all giant galaxies from the sample and deleted closely interacting and perturbed galaxies.  After all these selections, I had a sample of 272 giant discy galaxies.

 The giant galaxies of my sample have similar \HI masses as the most \HI  massive and fastest rotating disc galaxies from the sample of  \cite{Courtois2015} -- the median value of \HI mass for my sample is 2.36x$10^{10}$ solar masses. The stellar masses for the galaxies of my sample with available $(B-V)_0$ colour indices are comparable to that of the superluminous spiral galaxies from  \cite{Ogleetal2016}. The closest galaxy of my sample is at the distance of roughly 17 Mpc and the most distant is at 662 Mpc. My sample have 7 common objects with the sample of \cite{Courtois2015}, but none with that of  \cite{Ogleetal2016}, since it contains predominantly more distant galaxies than in the current sample. 

Most of the galaxies of the sample are not LSB, since for LSB galaxy the radius $R_{25}$ should be significantly lower than the disc size due to low central surface brightness, if not to propose that giant HSB disc is surrounded by even more giant LSB disc which seems to be very exotic. 

To learn more on the evolution of giant discy galaxies it is important to compare their properties to that of the entire sample of galaxies with the same types, inclinations, but without any restrictions on the disc sizes. So I made similar request to the database as for giant galaxies but without the specification for $R_{25}$.  The resulted entire sample consisted of 23756 systems. I used this sample for the comparison with the sample of giant discy galaxies. All properties that I used in the statistical analysis in the next section were taken from Hyperleda database. 
\section{Statistics}\label{Stat}
As on can expect giant discy galaxies appear to have higher B-band luminosities, than the entire sample (see Fig. \ref{hist_lb}). This difference is significant according to Kolmogorov-Smirnov (hereafter K-S) test with the significance level 0.05. The giant discy galaxies also have higher rotation velocities (see Fig. \ref{hist_vrot}), which is also significant as follows from K-S test. This difference indicates that giant discs are both more luminous and more massive than discy galaxies of moderate sizes. I calculated roughly the dynamical masses of the galaxies inside the optical borders using the following formula: 
{$M\approx V^2R_{25}/G$, where $R_{25}$ is the optical radius of object, $V$ is its rotation velocity corrected for inclination and G is gravitational constant. The dynamical mass-to-light ratio can give one clues on the fraction of the dark matter in galaxies, so I compared the galaxies by the ratios of the dynamical masses to B-band luminosities (see Fig. \ref{hist_mlb}). As one can see from the histograms, the giant discy galaxies tend to have higher dynamical mass-to-light ratios (the difference is significant according to K-S test). Partly this difference can be explained by higher gas masses of giant discy galaxies (see Fig. \ref{hist_mhi}). However the \HI mass to B-band luminosity ratios of giant discs are statistically not different from the entire sample if to delete the latest type galaxies from the entire sample, since they are absent in the sample of giants (see Fig. \ref{hist_type})\footnote{before the removing of the latest type galaxies from the total sample, giant galaxies tended to have even lower \HI mass to light ratios in comparison to the entire sample}. It is in good agreement with  \cite{Courtois2015} who found that  there is no difference in stellar-to-baryonic mass ratio between the \HI-massive sample they studied  and the typical disc galaxies. I also found that giant discy galaxies do not deviate from the correlation $\log(M_{\rm HI})(R_{25})$ found for the galaxies with moderate sizes (see Fig. \ref{mhir25}, where the line shows the fit for the moderate size galaxies and squares and circles correspond to giant and moderate size galaxies). It can give evidences of the absence of the peculiarities in the evolution of star formation of giant spirals. I can conclude that most of giant discs did not experienced very strong star formation bursts or strangulation. 

The giant discy galaxies differ from 'normal' ones by the total-to-baryonic mass ratio (the baryonic mass is calculated as a sum  of the \HI mass and stellar mass estimated from the $(B-V)_0$ colour indices and model M/L-colour relations from \cite{bdj}). The giant discy galaxies tend to have higher total-to-baryonic mass ratios in comparison to the entire sample, although the statistics for giant galaxies becomes poor due to the lack of colour indices for the bulk of these systems. Thus the difference between dynamical mass-to-light ratios  could be due to the higher fraction of dark matter either in baryonic (i.e. cold gas non-detected by its emission, see, e.g. \citealt{Kasparova2014}) or no-baryonic form in giant discs. 

Other specialties of giant discs are their shortage at the distances $D<100$ Mpc (see Fig. \ref{hist_d}) and their lower mean B-band surface brightnesses of the discs (see Fig. \ref{hist_bri25}). The latter can indicate that giant discs are more rarified in comparison to typical late type galaxies. 

Since bars can play important role in evolution of galaxies I decided to compare the samples by the fraction of barred galaxies. According to my analysis the bars are slightly more frequent for giant discs in comparison to moderate size galaxies. I found the following frequencies of bar occurrence: 1/2.67 and 1/3.2 for giant galaxies and the entire sample respectively. According to \cite{Seideletal2015} bars do not strongly influence the global kinematics of their host galaxies, however they can have noticeable effect on the inner kinematics inducing the formation of inner disc or a peanut-shaped bulge (see e.g. \citealt{Saburovaetal2017}). The bars in principle can fuel with gas the central black hole which mass in turn can be related to the processes of fading of star-formation in giant discs. \cite{Ogleetal2016} proposed that the insufficient mass of supermassive black hole can lead to the absence of strong feedback that drives away the gas from giant discs.  However it seems unlikely that bar has significant influence on the AGN strength (see e.g. \citealt{Cisternasetal2015}). Thus it is hardly probable that bars are strongly connected to the high disc sizes of giant galaxies. 

Interestingly the ring occurs in giant galaxies two times more frequently than in the entire sample, the frequencies of the ring are 1/4.69 and 1/10.2 for the giant discs and the entire sample respectively. It can in principle indicate that at least a part of these objects has something in common with the largest ring galaxy UGC7069 studied by \cite{Ghosh2008} (see Sect. \ref{intro}). At the same time the higher luminosities, rotation velocities, \HI masses of giant spirals and the fact that they follow the trend $\log(M_{\rm HI})(R_{25})$ found for normal size galaxies speak against the catastrophic scenario of the formation of these systems in which the normal-size galaxy is blown up. In this case one would expect them to have similar masses as moderate size spirals and to lie below the correlation $\log(M_{\rm HI})(R_{25})$ which contradicts the results of current paper. 

Another feature that could be important is the distribution of giant discs by inclination. In Fig.  \ref{hist_incl} I show the histograms plotted for the inclinations of giant galaxies and the entire sample. One can see that there is an excess of objects with $i=90$\degr~both for the entire sample and for giant galaxies. This excess is due to fact that some of objects are erroneously considered as edge-on galaxies. One can see that the fraction of the galaxies with $i=90$\degr~is higher for the sample of giants. The dust extinction has higher influence on the surface photometry of the systems in a positions close to edge-on. However, since I used B-band determination of the optical sizes I don't expect significant overestimation  of $R_{25}$ due to the dust extinction. 

Since the environment can have significant influence on the evolution of galaxies I paid attention to the frequencies of the membership of systems to associations such as groups and clusters and compared these frequencies for the giant discs to that for the total sample of galaxies.  To do it I utilized the Hierarchy catalogue from Hyperleda and the galaxy group catalogues based on the 2MRS and SDSS DR12 galaxy samples by \cite{Saulderetal2016}. According to both sources giant galaxies tend to be slightly more rarely in groups and in clusters. This conclusion is in good agreement with \cite{Courtois2015} who found that none of the giant disc galaxies from their sample is in a highly overdense environment. However, \cite{Courtois2015} also concluded that the most massive
\HI detected galaxies are located preferentially in filaments. These conclusions can indicate that it is environment at the stage of galaxies formation that had the highest impact on the evolution of giant systems and could aid in the formation of the large discs, but not the current environment.

\begin{figure}
\includegraphics[width=0.45\textwidth]{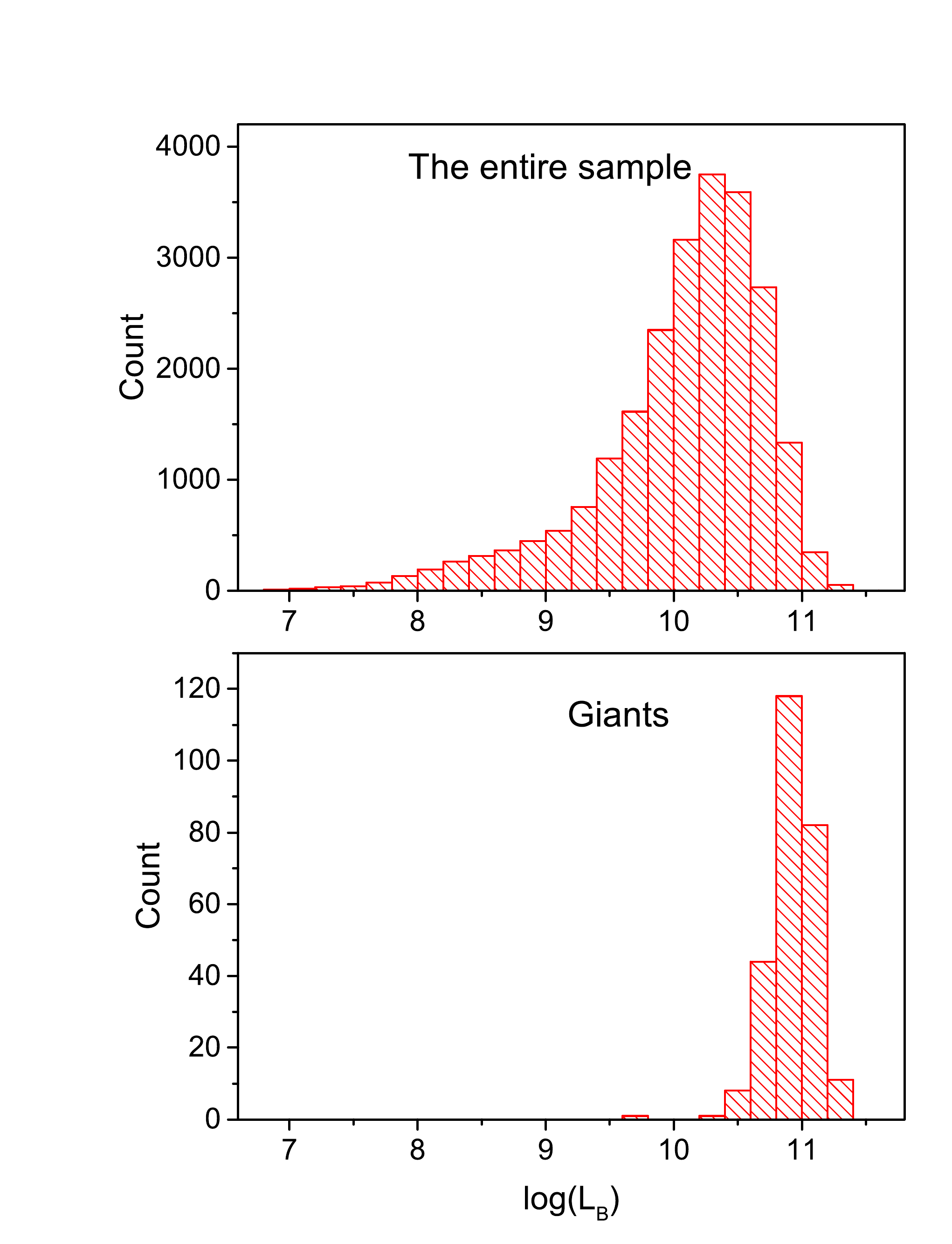}
\caption{The histograms for B-band luminosities of the entire sample (upper panel) and giant discy galaxies (bottom panel). The bins are the same for both histograms.}
\label{hist_lb} 
\end{figure}
\begin{figure}
\includegraphics[width=0.45\textwidth]{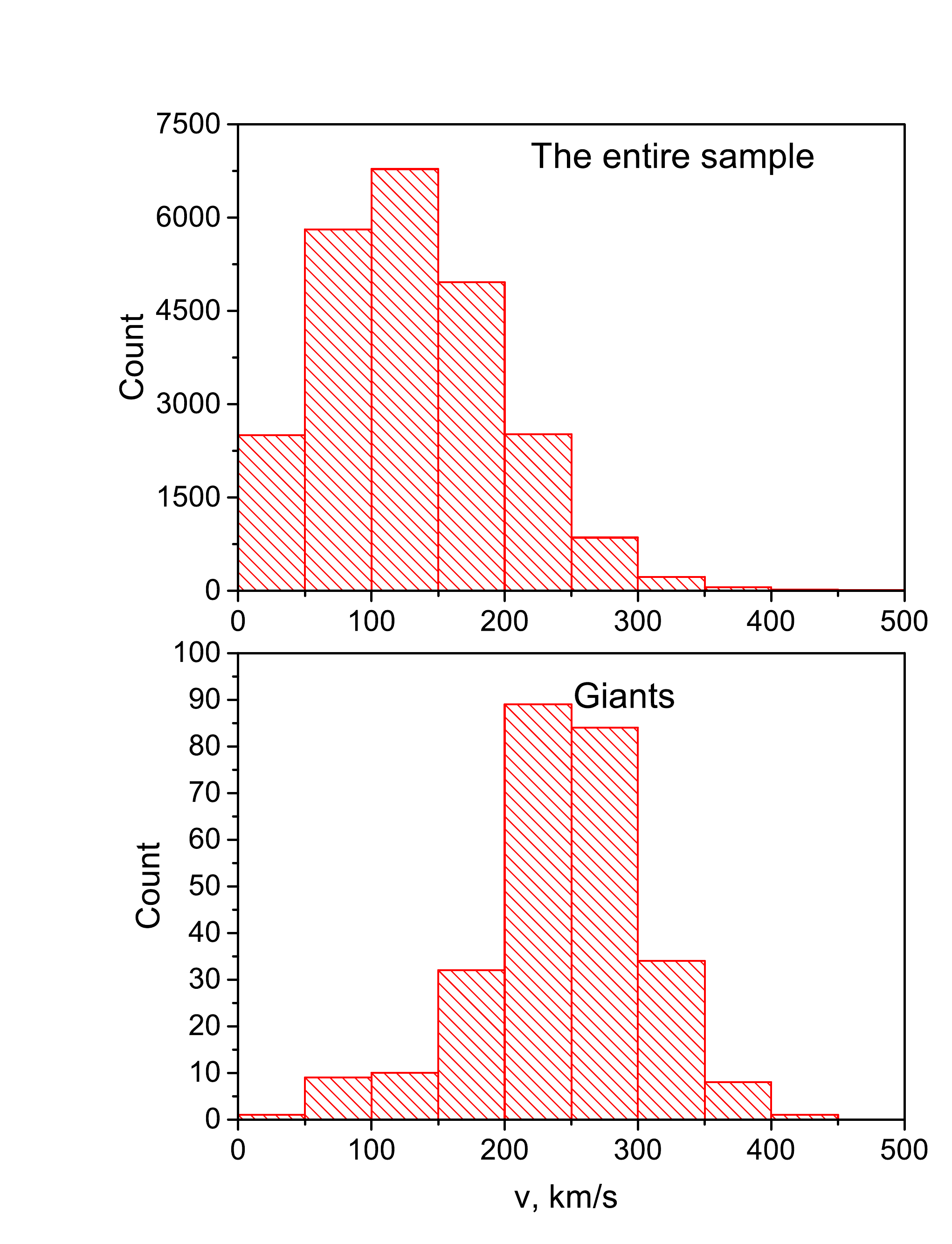}
\caption{The histograms for rotation velocity of the entire sample (upper panel) and giant discy galaxies (bottom panel). The bins are the same for both histograms. }
\label{hist_vrot} 
\end{figure}
\begin{figure}
\includegraphics[width=0.45\textwidth]{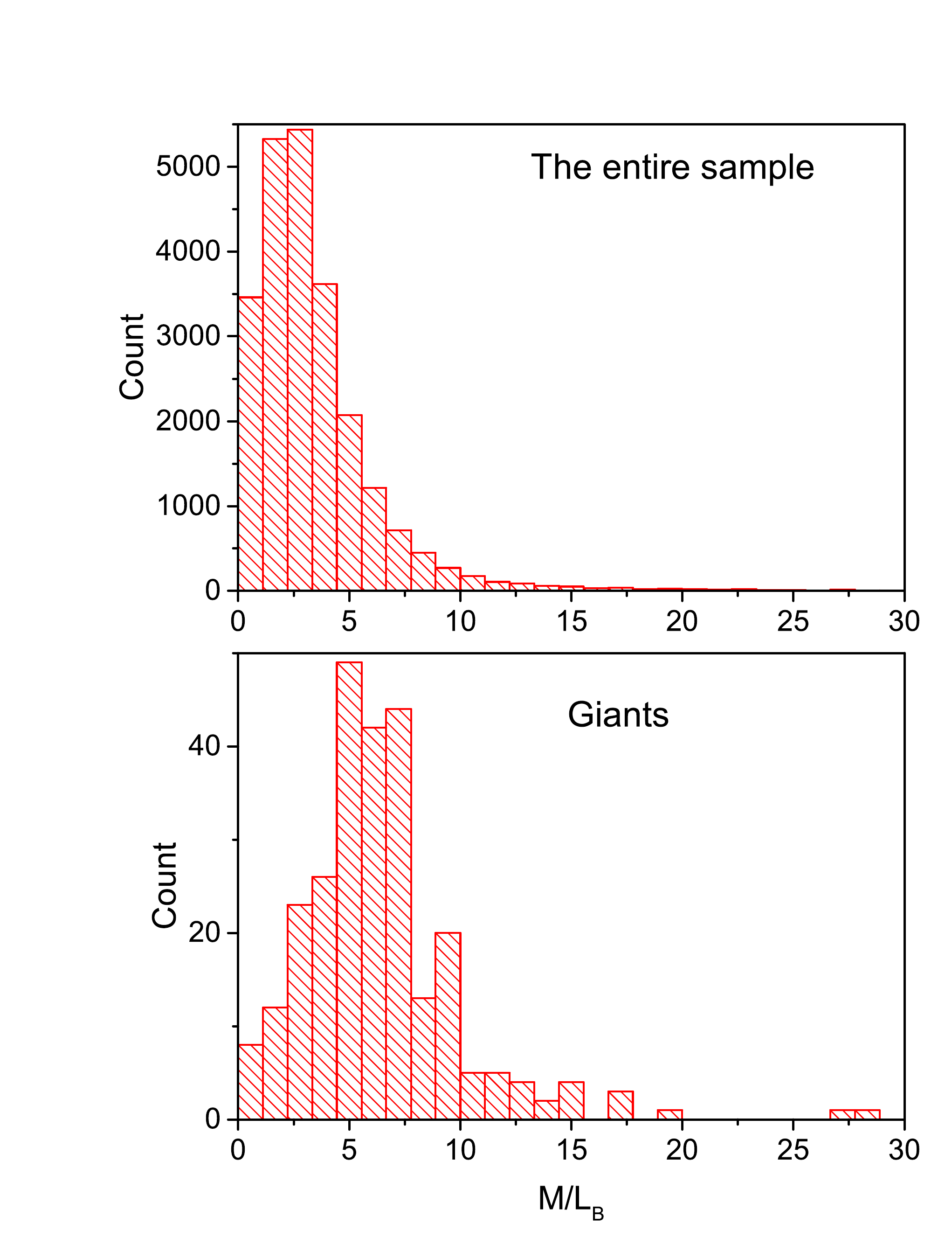}
\caption{The histograms for the dynamical mass to light ratio in B-band of the entire sample (upper panel) and giant discy galaxies (bottom panel). The bins are the same for both histograms.}
\label{hist_mlb} 
\end{figure}

\begin{figure}
\includegraphics[width=0.45\textwidth]{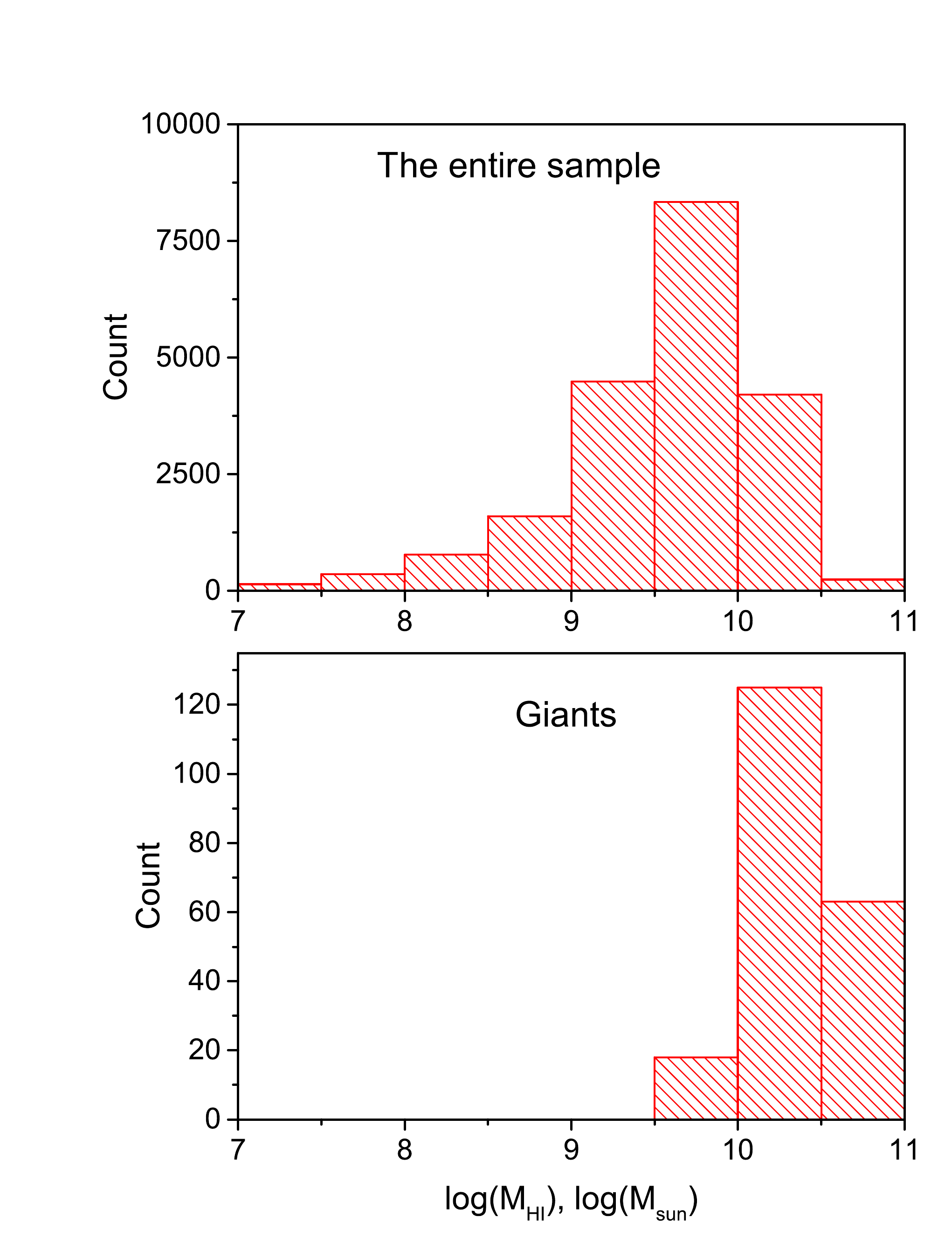}
\caption{The histograms for the \HI mass of the entire sample (upper panel) and giant discy galaxies (bottom panel). The bins are the same for both histograms.}
\label{hist_mhi} 
\end{figure}
\begin{figure}
\includegraphics[width=0.45\textwidth]{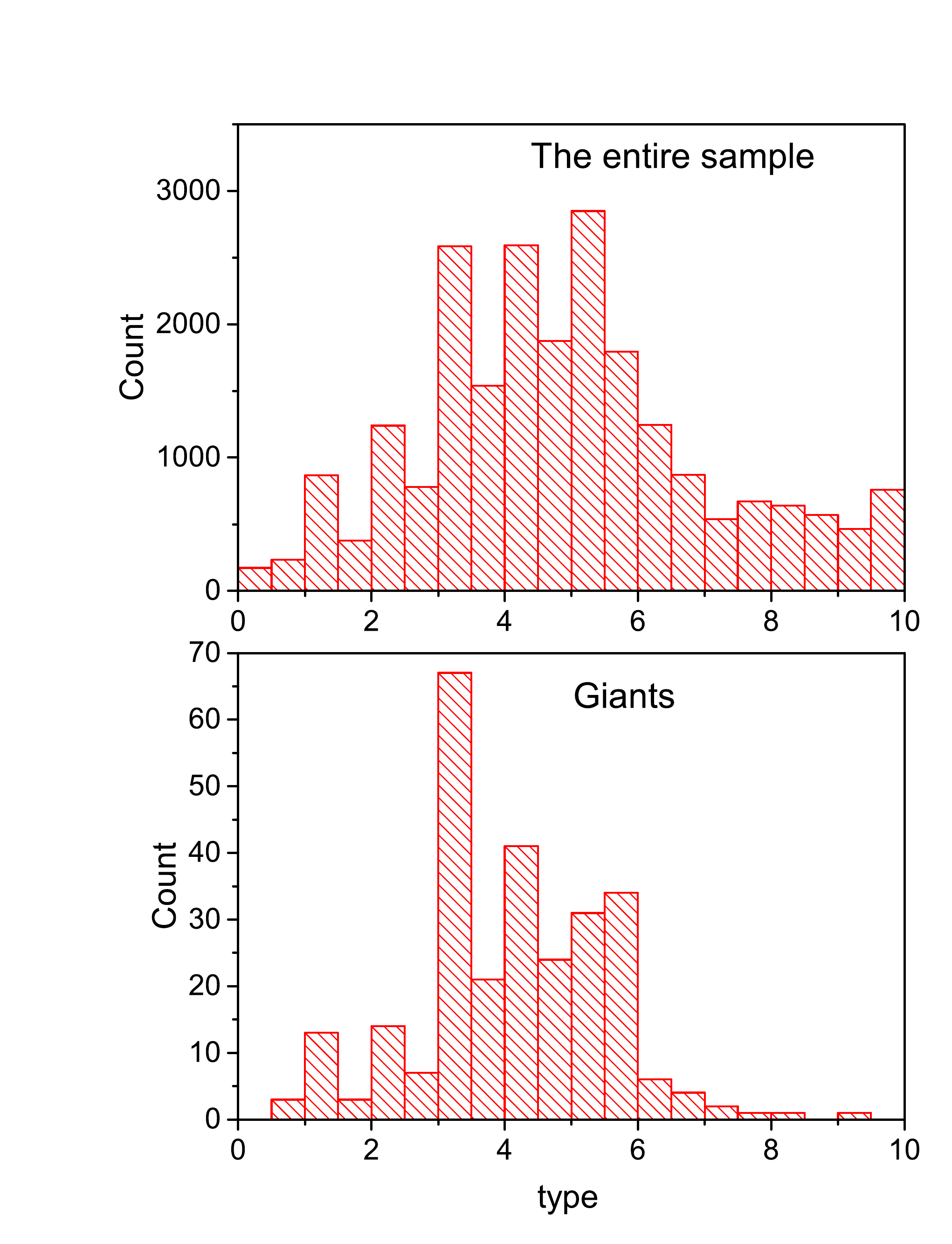}
\caption{The histograms for the morphological type of the entire sample (upper panel) and giant discy galaxies (bottom panel). The bins are the same for both histograms.}
\label{hist_type} 
\end{figure}

\begin{figure}
\includegraphics[width=0.45\textwidth]{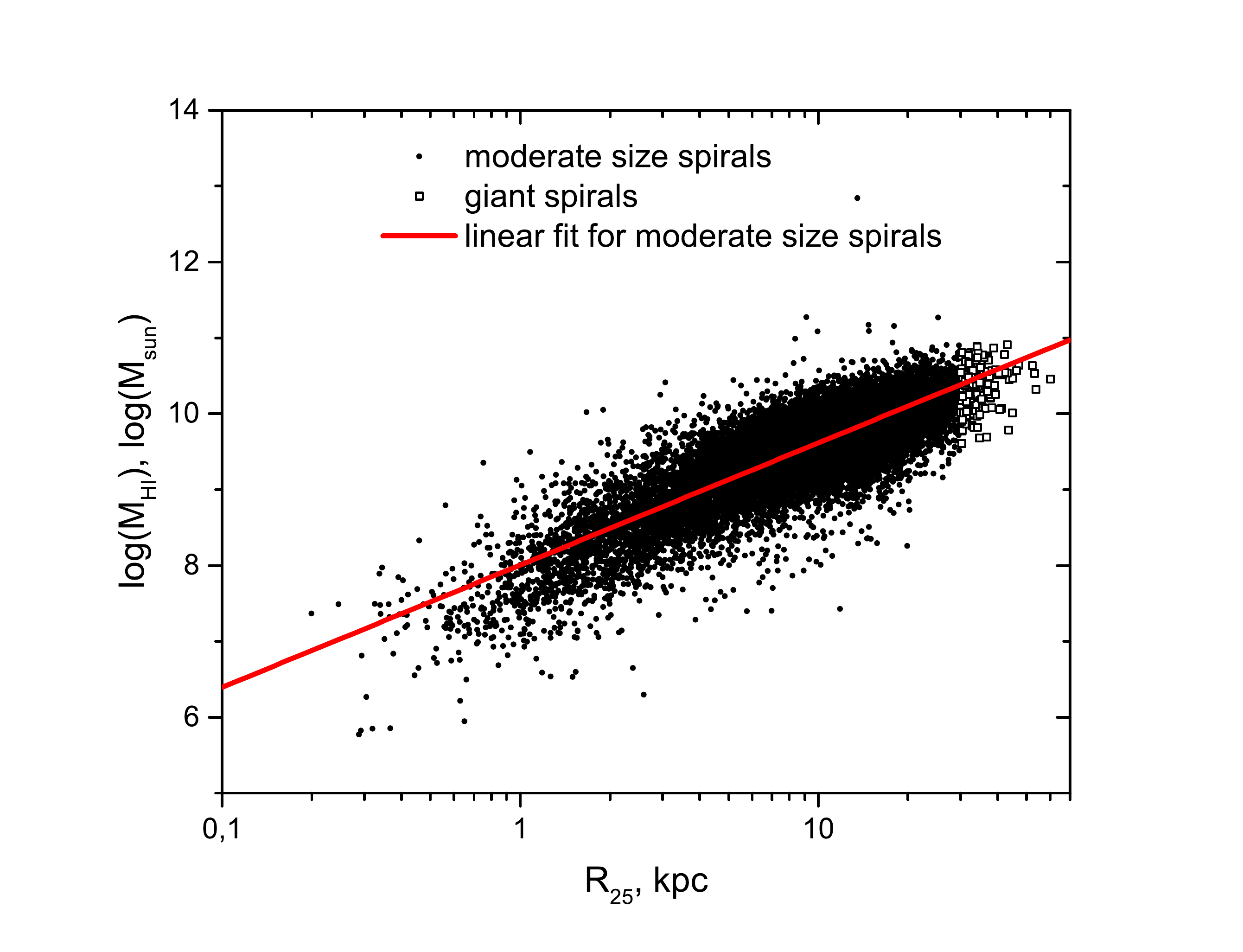}
\caption{The relation between the optical radius and \HI mass. Small circles show discy galaxies with $R_{25}<30$ kpc. Squares correspond to giant spirals. Line shows the fit for the galaxies with $R_{25}<30$ kpc.}
\label{mhir25} 
\end{figure}

\begin{figure}
\includegraphics[width=0.45\textwidth]{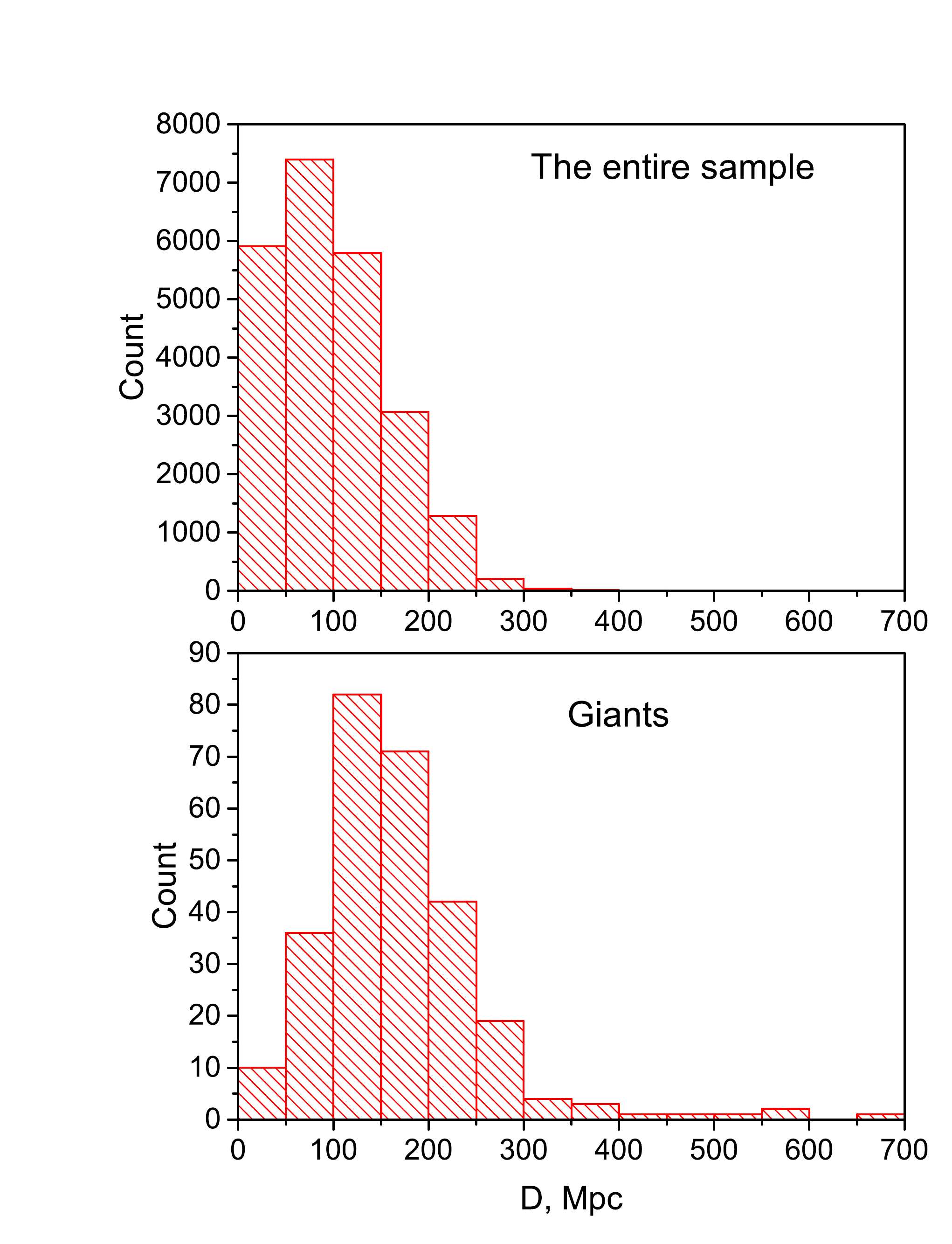}
\caption{The histograms for the distances of the entire sample (upper panel) and giant discy galaxies (bottom panel). The bins are the same for both histograms.}
\label{hist_d} 
\end{figure}

\begin{figure}
\includegraphics[width=0.45\textwidth]{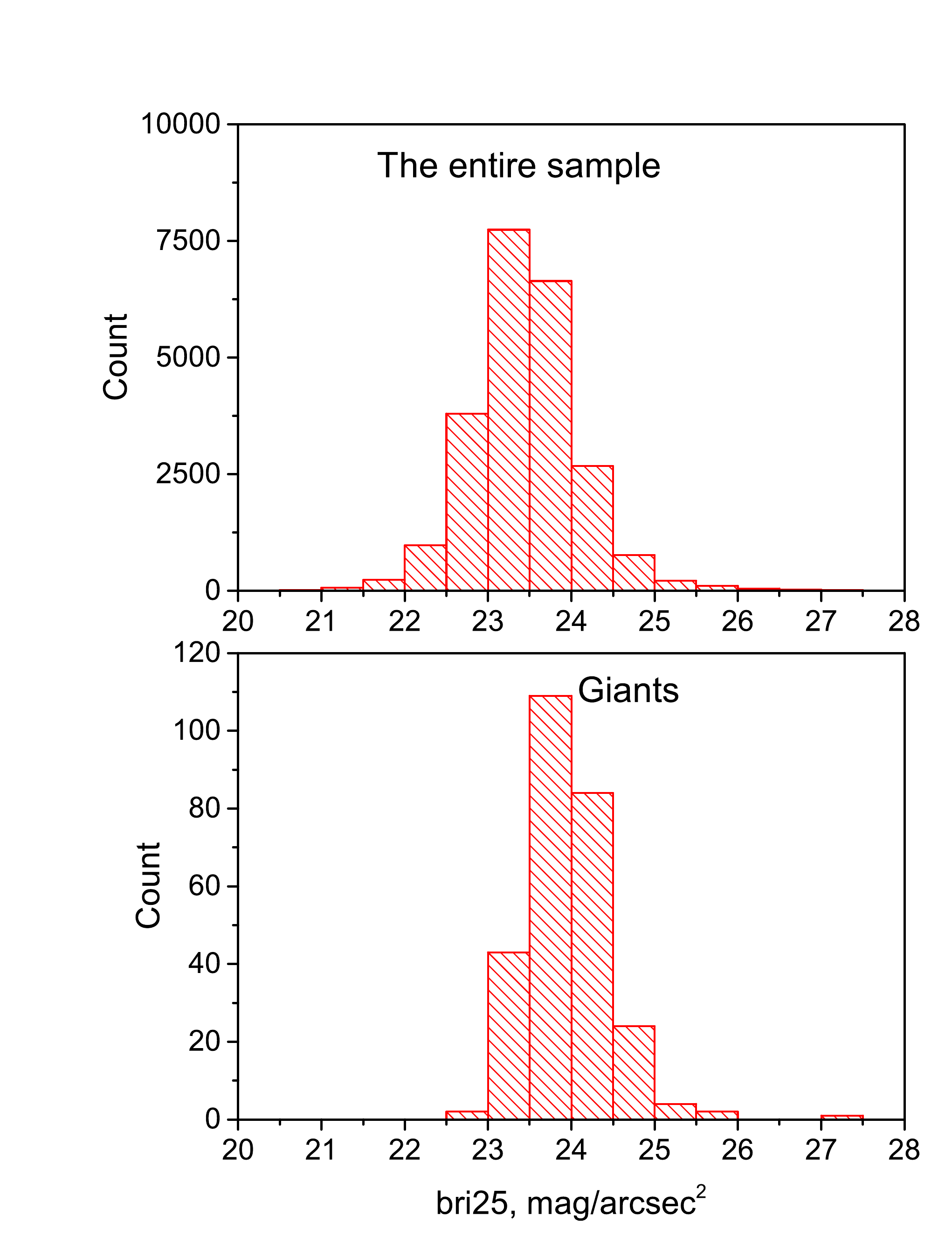}
\caption{The histograms for the 	mean surface brightness within B-band isophote 25 $mag/arcsec^2$ of the entire sample (upper panel) and giant discy galaxies (bottom panel). The bins are the same for both histograms.}
\label{hist_bri25} 
\end{figure}

\begin{figure}
\includegraphics[width=0.45\textwidth]{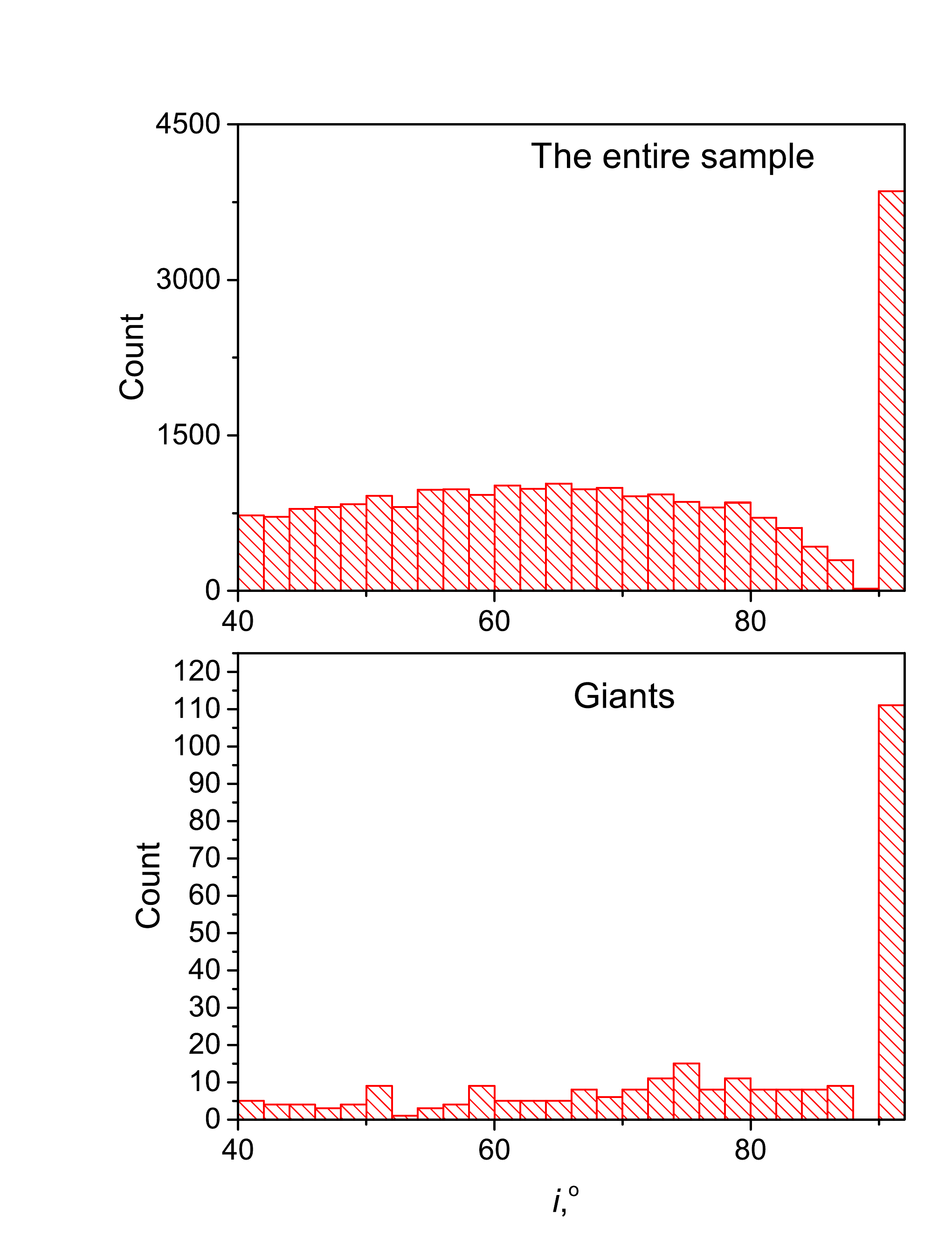}
\caption{The histograms for the 	inclination of the entire sample (upper panel) and giant discy galaxies (bottom panel). The bins are the same for both histograms.}
\label{hist_incl} 
\end{figure}

\section{On the parameters of dark haloes}\label{Parameters}
As far as dark matter (DM) halo plays important role in the evolution of  galaxies it is essential to know if there are any peculiarities in the dark matter haloes of giant discy galaxies. Thus, as the next step I referred to the compilation of Spatially Resolved Kinematics of galaxies\footnote{http://leda.univ-lyon1.fr/a005/} -- one of the Hyperleda projects and searched for the rotation curves of the giant galaxies. My goal was to obtain the dark matter halo parameters of the galaxies as reliably as it is possible using the decomposition of the rotation curve into the components. This method has its weak points, which are discussed e.g. in \cite{saburovaetal2016} and taken into account in current paper. Along with the rotation curve I required the photometrical parameters of discs and bulges. Eventually I had a sample of 18 giant galaxies with available rotation curves of satisfactory quality and the parameters of disc and bulge. In Table \ref{Tab1} I list the galaxies, the adopted distances, inclinations and the sources of the rotation curve and photometrical parameters. 

For some of the galaxies only radial surface brightness profiles were available but not the parameters of disc and bulge, for these systems I performed the decomposition of  profiles into the components of bulge and disc. For a bulge I utilized the Sersic profile of~surface brightness \citep{Sersic68}:
\begin{equation}
\label{eq:Sersic}
I_b(r) = 
          (I_0)_b 10^{\left[ - b_n\left( \frac{r}{R_{e}} \right)^{1/n} \right]}. \end{equation}
Here $(I_0)_b$ is the bulge central surface brightness, $R_e$ is the effective radius containing a half of the luminosity,
$b_n\approx 1.9992 n - 0.3271$ (\citealt{Caon1993})
and $n$ is the Sersic index. For the disc, I choose exponential radial distribution law: 
\begin{equation}I_d(r)=(I_{d})_{0}\exp(-r/R_d), \end{equation} where $(I_{d})_{0}$ and $R_d$ are the disc central surface brightness and the exponential scalelength, correspondingly.  I give the obtained parameters of discs and bulges in Table \ref{Tab2}. The parameters correspond to I-band for all galaxies except UGC02885 for which I used $3.6\mu \mathrm{m}$  data. 

\begin{table*}
\caption{The properties of giant discy galaxies with available rotation curves: adopted distance; inclination, the source papers for rotation curve and surface photometry  }\label{Tab1}
\begin{center}
\begin{tabular}{ccccc}
\hline\hline
Galaxy&D&$i$&Source of & Source of\\
    & (Mpc)     & (\degr) &  rotation curve     & bulge and disc parameters \\
\hline
ESO030-009&114.8&82.1&\cite{Mathewson1992}&\cite{Mathewson1992}\\
ESO184-051&103&78.4&\cite{Mathewson1992}&\cite{Mathewson1992}\\
ESO240-011&35.4&90&\cite{Kregeletal2004}&\cite{Kregeletal2002}\\
ESO382-058&106.2&79&\cite{Mathewson1992}&\cite{Palunas2000}\\
ESO460-031&84&90&\cite{Mathewson1992}&\cite{Mathewson1992}\\
ESO471-002&121&73.2&\cite{Mathewson1992}&\cite{Mathewson1992}\\
ESO563-021&70.1&83&\cite{Spekkens2006}&\cite{Spekkens2006}\\
IC0724&125.6&73.1&\cite{Corsinietal1999}&\cite{Moriondo1998}\\
IC1862&89.2&90&\cite{Mathewson1992}&\cite{Mathewson1992}\\
IC2531&36.8&86-90&\cite{Allaertetal2015}&\cite{Xilouris1999}\\
NGC0973&63.5&88&\cite{Allaertetal2015}&\cite{Xilouris1997}\\
NGC4705&62.9&78.4&\cite{Mathewson1992}&\cite{Mathewson1992}\\
NGC5529&49.5&85-87.5&\cite{Allaertetal2015}&\cite{Xilouris1999}\\
NGC7184&34.67&90&\cite{Mathewson1992}&\cite{Mathewson1992}\\
PGC006966&70.2&90&\cite{Mathewson1992}&\cite{Mathewson1992}\\
UGC02885&80.6&72&\cite{Struveetal2007}&\cite{Lellietal2016}\\
UGC04277&76.5&87.3&\cite{Allaertetal2015}&\cite{Bizyaev2002}\\
UGC05341&99.85&90&
\cite{Catinella2005}&\cite{Bizyaev2002}\\
\hline\hline
\end{tabular}
\end{center}
\end{table*}

\begin{table*}
\caption{The photometrical parameters of discs and bulges of the giant discy galaxies obtained in current paper: central surface brightness of bulge; effective radius of bulge; Sersic index of bulge; central surface brightness of exponential disc; radial scalelength of disc  }\label{Tab2}
\begin{center}
\begin{tabular}{cccccc}
\hline\hline
Galaxy&$(\mu_0)_b$&$R_e$&n&$(\mu_0)_d$&$R_d$\\
&($mag/arcsec^2$)&(arcsec)&&($mag/arcsec^2$)&(arcsec)\\ 
\hline
ESO030-009&18.23&4.6&0.74&19.59&23.27 \\
ESO184-051&15.86&2.26&1.56&19.21&17.02 \\
ESO382-058&15.17&0.71&1.45&19.82&9.66 \\
ESO460-031&18.27&8.81&1.02&20.18&25.98 \\
ESO471-002&17.95&4.09&1.61&19.42&12.59 \\
ESO563-021&17.86&1.65&0.80&19.04&8.37 \\
IC1862&18.38&4.41&0.77&19.11&17.09 \\
NGC4705&17.40&16.33&1.63&19.85&29.06\\
NGC7184&17.42&5.48&0.69&18.88&40.37\\
PGC006966&18.53&5.26&0.46&19.24&20.75\\
UGC02885&13.38&7.15&1.98&17.78&26.57\\
\hline\hline
\end{tabular}
\end{center}
\end{table*}

I decomposed the collected rotation curves into the components of gaseous and stellar discs, bulge and dark halo. To do it I used the approach described in details in \cite{saburovaetal2016}, its basic concept is to calculate the $\chi^2$ for the grid of values of the DM halo parameters.  For the dark matter haloes I used the following density profiles.  The density profile by \citet{Burkert1995}: 
\begin{equation}\label{Burkert}
\rho_{\mathrm{burk}}(r)=\frac{\rho_0 R_s^3}{(r+R_s)(r^2+R_s^2)}. 
\end{equation} 
Here $\rho_{0}$ and $R_{s}$ are the central density and the radial scale of the halo\footnote{Below $R_s$ and $\rho_0$ are different for the various DM density profiles.}.  
 The pseudoisothermal profile (hereafter, piso):  
\begin{equation}    
\rho_{\mathrm{piso}}(r)=\frac{\rho_{0}}{(1+(r/R_{s})^2)},
\end{equation}
(The Navarro-Frenk-White profile \citet{nfw1996}  (hereafter, NFW): \begin{equation}    
\rho_{\mathrm{nfw}}(r)=\frac{\rho_{0}}{(r/R_s)(1+(r/R_{s })^2)^{2}},
\end{equation} 
 
I took the gaseous surface density radial profiles from the sources of the rotation curves if they were available. In the opposite case I fixed the gas density at zero level. The uncertainties of the rotation velocity amplitude were given by far not in every paper, which is a separate problem. In cases of the absence of error bars I used the mean dispersion of the points as the error. I  made tests with different values of errors and concluded that it has no significant influence on the main conclusion of current paper but it can affect the uncertainties of the dark halo central density and radial scale, which are known to be ambiguous  due to the degeneracy of these parameters and are already taken with great caution (see below).

 In order to decrease the uncertainties of the parameters of dark matter haloes I decided to fix the mass-to-light ratios of disc and bulge in narrow ranges followed from the population synthesis modeling and the observed colour indices. I used $M/L$-colour relations from \cite{McGaughSchombert2014} for different stellar initial mass functions to get the lower and upper limits of mass-to-light ratio.

For several galaxies I got the infinite contours  of  $1\sigma$ confidence levels on the $\chi^2$ map.   It means that it is impossible to obtain the trustworthy estimates of the central density and radial scale of the dark halo due to the degeneracy between these parameters, however the estimate of halo mass inside of optical radius is almost not affected by the degeneracy (see \citealt{saburovaetal2016}). In Figs.\ref{map_a} and \ref{map_b} I show the examples of  the $\chi^2$ maps for UGC04277 (closed $1 \sigma$ contours) and ESO 240-011 (infinite contours  of  $1\sigma$) respectively. Left panels correspond to the models with the piso profiles of the dark halo, centre panels~--- to the Burkert  profiles, right panels~--- to the NFW profiles.
The colour on the maps denotes the $\chi^2$ value, the redder the colour, the lower the $\chi^2$ and the better is the fit. 
The black contours refer to $1\sigma$, $2\sigma$ and $3\sigma$ confidence limits. 
 The lower row for each example gives the best-fitting decomposition
corresponding to the $\chi^2$  minimum, where the black symbols  mark the observed rotation curve, thick red line~--- the total model, cyan  line~---  dark halo, blue line~--- stellar disc, magenta line~---  gas disc, green line~---  bulge.
In Tables ~\ref{parameters_burk}--\ref{parameters_nfw}  I give the resulting parameters corresponding to the minimal values of $\chi^2$ for each type of DM density profile  together with the errors associated with the range covered by $1\sigma$ confidence limit (each pair of the parameters of DM halo on the map corresponds to the certain values of $\chi^2$, disc mass-to-light ratio and bulge central surface density). 
   I discuss the results for each galaxy below.

\begin{table*}
\begin{center}
\caption{The obtained parameters of the main components of the galaxies for Burkert dark halo density profile. 
The errors correspond to $1\sigma$ confidence limit. For galaxies with infinite $1\sigma$ contour (see the discussion in Sect.\ref{notes}) I put 0.00 as error. The columns contain the following data:
(1)~-- galaxy name;
(2) and (3)~-- radial scale and central density of the DM halo;
(4)~-- optical radius (4 disc scalelengths);
(5)~-- mass of DM halo inside of optical radius;
(6)~-- disc mass-to-light ratio, the band is noted in Sect. \ref{notes} for each galaxy;
(7)~-- central surface density of bulge\label{parameters_burk};
}
\renewcommand{\arraystretch}{1.5}
\begin{tabular}{lrlrl r rlrlrl}
\hline
Galaxy	&	\multicolumn{2}{c}{$R_s$}&	\multicolumn{2}{c}{$\rho_0$ }&	\multicolumn{1}{c}{$R_{opt}$}	&	\multicolumn{2}{c}{$M_{halo}$}	 &	\multicolumn{2}{c}{$M/L$} &	\multicolumn{2}{c}{$(I_0)_b$}\\
&\multicolumn{2}{c}{kpc}&\multicolumn{2}{c}{$10^{-3}$ M$_{\odot}/$pc$^3$}& \multicolumn{1}{c}{kpc}&\multicolumn{2}{c}{$10^{10}$ M$_{\odot}$}&\multicolumn{2}{c}{M$_{\odot}/$L$_{\odot}$	}& \multicolumn{2}{c}{$10^{3}$ M$_{\odot}/$pc$^2$}\\
\hline
\hline
ESO030-009&    200.00& $^{+      0.00}_{-      0.00} $  &      0.50& $^{+      0.00}_{-      0.00} $  &     51.78&     23.25& $^{+      1.56}_{-      1.69} $  &      2.54&&      1.80&\\
ESO184-051&      5.44& $^{+      2.38}_{-      1.17} $  &    130.29& $^{+     95.87}_{-    100.77} $  &     34.05&     31.89& $^{+      7.06}_{-     20.61} $  &      0.23&&      3.22&\\
ESO240-011&    100.00& $^{+      0.00}_{-      0.00} $  &      1.72& $^{+      0.00}_{-      0.00} $  &     27.47&     11.86& $^{+      2.79}_{-      1.50} $  &      1.58&&    864.30&\\
ESO382-058&      9.13& $^{+      5.16}_{-      2.33} $  &     57.84& $^{+     44.94}_{-     37.17} $  &     38.64&     49.37& $^{+      7.76}_{-     15.12} $  &      1.90&&     29.21&\\
ESO460-031&    100.00& $^{+      0.00}_{-      0.00} $  &      1.21& $^{+      0.00}_{-      0.00} $  &     42.32&     26.44& $^{+     10.05}_{-     12.82} $  &      1.68&&      1.49&\\
ESO471-002&    200.00& $^{+      0.00}_{-      0.00} $  &      0.97& $^{+      0.00}_{-      0.00} $  &     29.52&      9.32& $^{+      7.95}_{-      8.69} $  &      1.95&&      2.78&\\
ESO563-021&     20.33& $^{+      6.48}_{-      3.56} $  &     12.57& $^{+      5.11}_{-      5.13} $  &     34.10&     41.18& $^{+      2.57}_{-      5.76} $  &      1.46&&      1.12&\\
IC0724&    700.00& $^{+      0.00}_{-      0.00} $  &      0.97& $^{+      0.00}_{-      0.00} $  &     45.76&     37.16& $^{+      0.69}_{-      0.68} $  &      0.48&&     36.72&\\
IC1862&    700.00& $^{+      0.00}_{-      0.00} $  &      0.83& $^{+      0.00}_{-      0.00} $  &     29.56&      8.65& $^{+      5.07}_{-      4.96} $  &      1.19&&      0.76&\\
IC2531&     93.72& $^{+    134.96}_{-     22.74} $  &      1.19& $^{+      0.17}_{-      0.29} $  &     34.93&     15.36& $^{+      0.45}_{-      1.23} $  &      0.64&&   1705.76&\\
NGC0973&    200.00& $^{+      0.00}_{-      0.00} $  &      0.63& $^{+      0.00}_{-      0.00} $  &     44.35&     19.24& $^{+      2.03}_{-      3.68} $  &      0.71&&   3419.67&\\
NGC4705&     35.11& $^{+     46.00}_{-     12.01} $  &      2.04& $^{+      1.12}_{-      0.88} $  &     35.44&     14.33& $^{+      3.14}_{-      2.97} $  &      1.00&&      3.57&\\
NGC5529&     58.21& $^{+     16.07}_{-      8.98} $  &      2.13& $^{+      0.45}_{-      0.45} $  &     28.76&     13.69& $^{+      1.44}_{-      1.72} $  &      0.49&&    436.10&\\
NGC7184&     20.16& $^{+      0.00}_{-      0.00} $  &     12.32& $^{+      0.00}_{-      0.00} $  &     27.14&     27.75& $^{+     22.56}_{-     10.95} $  &      1.29&&      2.51&\\
PGC006966&     24.89& $^{+     15.29}_{-      7.17} $  &     12.49& $^{+      5.98}_{-      4.34} $  &     28.25&     39.14& $^{+      6.83}_{-      6.25} $  &      1.00&&      0.00&\\
UGC02885&     39.03& $^{+     15.41}_{-      9.30} $  &      3.78& $^{+      2.00}_{-      1.22} $  &     41.53&     40.54& $^{+      4.36}_{-      3.87} $  &      0.60&&     22.15&\\
UGC04277&     22.02& $^{+      2.03}_{-      2.63} $  &      9.21& $^{+      2.72}_{-      1.35} $  &     32.12&     30.88& $^{+      2.50}_{-      1.22} $  &      0.80&&   1256.11&\\
UGC05341&     26.19& $^{+      5.21}_{-      2.44} $  &     11.56& $^{+      1.66}_{-      2.73} $  &     35.28&     57.18& $^{+      2.11}_{-      3.58} $  &      0.37&&      0.00&\\
\hline
\end{tabular}
\end{center}
\end{table*}

\begin{table*}
\begin{center}
\caption{The obtained parameters of the main components of the galaxies for piso dark halo density profile. 
The errors correspond to $1\sigma$ confidence limit. For galaxies with infinite $1\sigma$ contour (see the discussion in Sect.\ref{notes}) I put 0.00 as error. The columns contain the following data:
(1)~-- galaxy name;
(2) and (3)~-- radial scale and central density of the DM halo;
(4)~-- optical radius (4 disc scalelengths);
(5)~-- mass of DM halo inside of optical radius;
(6)~-- disc mass-to-light ratio, the band is noted in Sect. \ref{notes} for each galaxy;
(7)~-- central surface density of bulge\label{parameters_piso};
}
\renewcommand{\arraystretch}{1.5}
\begin{tabular}{lrlrl r rlrlrl}
\hline
Galaxy	&	\multicolumn{2}{c}{$R_s$}&	\multicolumn{2}{c}{$\rho_0$ }&	\multicolumn{1}{c}{$R_{opt}$}	&	\multicolumn{2}{c}{$M_{halo}$}	 &	\multicolumn{2}{c}{$M/L$} &	\multicolumn{2}{c}{$(I_0)_b$}\\
&\multicolumn{2}{c}{kpc}&\multicolumn{2}{c}{$10^{-3}$ M$_{\odot}/$pc$^3$}& \multicolumn{1}{c}{kpc}&\multicolumn{2}{c}{$10^{10}$ M$_{\odot}$}&\multicolumn{2}{c}{M$_{\odot}/$L$_{\odot}$	}& \multicolumn{2}{c}{$10^{3}$ M$_{\odot}/$pc$^2$}\\
\hline
\hline
ESO030-009&    200.00& $^{+      0.00}_{-      0.00} $  &      0.44& $^{+      0.00}_{-      0.00} $  &     51.78&     24.32& $^{+      1.83}_{-      1.59} $  &      2.55&&      1.80&\\
ESO184-051&      2.48& $^{+      2.37}_{-      1.35} $  &     75.85& $^{+    258.59}_{-     52.29} $  &     34.05&     17.80& $^{+     18.93}_{-      3.85} $  &      1.00&&      3.22&\\
ESO240-011&    100.00& $^{+      0.00}_{-      0.00} $  &      1.39& $^{+      0.00}_{-      0.00} $  &     27.47&     11.56& $^{+      2.79}_{-      1.46} $  &      1.59&&    856.21&\\
ESO382-058&      4.51& $^{+      3.97}_{-      1.69} $  &     65.92& $^{+     82.68}_{-     46.76} $  &     38.64&     54.12& $^{+      8.15}_{-     16.64} $  &      1.90&&     29.21&\\
ESO460-031&    100.00& $^{+      0.00}_{-      0.00} $  &      1.00& $^{+      0.00}_{-      0.00} $  &     42.32&     28.64& $^{+     11.41}_{-     14.24} $  &      1.68&&      1.49&\\
ESO471-002&    200.00& $^{+      0.00}_{-      0.00} $  &      0.88& $^{+      0.00}_{-      0.00} $  &     29.52&      9.37& $^{+      8.18}_{-      8.70} $  &      1.97&&      2.74&\\
ESO563-021&     10.92& $^{+      4.19}_{-      2.52} $  &     13.46& $^{+      7.04}_{-      5.84} $  &     34.10&     41.00& $^{+      2.16}_{-      5.15} $  &      1.46&&      1.12&\\
IC0724&    700.00& $^{+      0.00}_{-      0.00} $  &      0.97& $^{+      0.00}_{-      0.00} $  &     45.76&     38.73& $^{+      0.73}_{-      0.76} $  &      0.48&&     36.72&\\
IC1862&      3.76& $^{+      0.00}_{-      0.00} $  &     22.35& $^{+      0.00}_{-      0.00} $  &     29.56&      9.56& $^{+      4.21}_{-      5.40} $  &      0.94&&      0.76&\\
IC2531&     58.84& $^{+     39.91}_{-     14.30} $  &      1.05& $^{+      0.16}_{-      0.18} $  &     34.93&     15.50& $^{+      0.66}_{-      1.08} $  &      0.64&&   1725.73&\\
NGC0973&    100.00& $^{+      0.00}_{-      0.00} $  &      0.59& $^{+      0.00}_{-      0.00} $  &     44.35&     19.30& $^{+      2.36}_{-      3.38} $  &      0.71&&   3415.19&\\
NGC4705&     23.10& $^{+     26.67}_{-      8.93} $  &      1.78& $^{+      1.04}_{-      0.72} $  &     35.44&     14.93& $^{+      2.73}_{-      3.05} $  &      1.00&&      3.57&\\
NGC5529&     37.52& $^{+     11.54}_{-      6.62} $  &      1.87& $^{+      0.41}_{-      0.43} $  &     28.76&     13.95& $^{+      1.44}_{-      2.00} $  &      0.49&&    441.16&\\
NGC7184&     12.35& $^{+      0.00}_{-      0.00} $  &     11.38& $^{+      0.00}_{-      0.00} $  &     27.14&     28.37& $^{+     22.24}_{-     10.87} $  &      1.29&&      2.51&\\
PGC006966&     15.65& $^{+      8.48}_{-      4.16} $  &     11.04& $^{+      5.06}_{-      3.73} $  &     28.25&     39.33& $^{+      7.25}_{-      5.02} $  &      1.00&&      0.00&\\
UGC02885&     22.57& $^{+      8.69}_{-     20.98} $  &      3.66& $^{+    383.72}_{-      1.19} $  &     41.53&     40.51& $^{+     10.92}_{-      2.72} $  &      0.60&&     22.15&\\
UGC04277&      4.93& $^{+      7.37}_{-      1.24} $  &     51.75& $^{+     37.27}_{-     42.63} $  &     32.12&     39.72& $^{+      1.64}_{-     10.04} $  &      0.40&&   1197.98&\\
UGC05341&     16.22& $^{+      3.68}_{-      1.85} $  &     10.46& $^{+      1.72}_{-      2.64} $  &     35.28&     58.02& $^{+      2.48}_{-      3.51} $  &      0.37&&      0.00&\\

\hline
\end{tabular}
\end{center}
\end{table*}

\begin{table*}
\begin{center}
\caption{The obtained parameters of the main components of the galaxies for NFW dark halo density profile. 
The errors correspond to $1\sigma$ confidence limit. For galaxies with infinite $1\sigma$ contour (see the discussion in Sect.\ref{notes}) I put 0.00 as error. The columns contain the following data:
(1)~-- galaxy name;
(2) and (3)~-- radial scale and central density of the DM halo;
(4)~-- optical radius (4 disc scalelengths);
(5)~-- mass of DM halo inside of optical radius;
(6)~-- disc mass-to-light ratio, the band is noted in Sect. \ref{notes} for each galaxy;
(7)~-- central surface density of bulge\label{parameters_nfw};
}
\renewcommand{\arraystretch}{1.5}
\begin{tabular}{lrlrl r rlrlrl}
\hline
Galaxy	&	\multicolumn{2}{c}{$R_s$}&	\multicolumn{2}{c}{$\rho_0$ }&	\multicolumn{1}{c}{$R_{opt}$}	&	\multicolumn{2}{c}{$M_{halo}$}	 &	\multicolumn{2}{c}{$M/L$} &	\multicolumn{2}{c}{$(I_0)_b$}\\
&\multicolumn{2}{c}{kpc}&\multicolumn{2}{c}{$10^{-3}$ M$_{\odot}/$pc$^3$}& \multicolumn{1}{c}{kpc}&\multicolumn{2}{c}{$10^{10}$ M$_{\odot}$}&\multicolumn{2}{c}{M$_{\odot}/$L$_{\odot}$	}& \multicolumn{2}{c}{$10^{3}$ M$_{\odot}/$pc$^2$}\\
\hline
\hline
ESO030-009&    200.00& $^{+      0.00}_{-      0.00} $  &      0.09& $^{+      0.00}_{-      0.00} $  &     51.78&     22.15& $^{+      2.06}_{-      2.15} $  &      2.39&&      1.80&\\
ESO184-051&     13.77& $^{+     24.29}_{-      7.60} $  &      9.75& $^{+     34.41}_{-      7.98} $  &     34.05&     17.03& $^{+     11.77}_{-      4.71} $  &      1.00&&      3.22&\\
ESO240-011&    100.00& $^{+      0.00}_{-      0.00} $  &      0.58& $^{+      0.00}_{-      0.00} $  &     27.47&     19.69& $^{+      1.91}_{-      3.13} $  &      1.21&&    862.62&\\
ESO382-058&     41.68& $^{+    197.22}_{-     24.27} $  &      2.97& $^{+     12.20}_{-      2.68} $  &     38.64&     47.18& $^{+     13.51}_{-     10.14} $  &      2.45&&     29.21&\\
ESO460-031&    100.00& $^{+      0.00}_{-      0.00} $  &      0.22& $^{+      0.00}_{-      0.00} $  &     42.32&     15.59& $^{+      6.78}_{-      9.05} $  &      1.68&&      1.49&\\
ESO471-002&    200.00& $^{+      0.00}_{-      0.00} $  &      0.11& $^{+      0.00}_{-      0.00} $  &     29.52&      9.85& $^{+      4.50}_{-      9.31} $  &      1.80&&      2.56&\\
ESO563-021&    228.06& $^{+      0.00}_{-      0.00} $  &      0.24& $^{+      0.00}_{-      0.00} $  &     34.10&     33.74& $^{+      6.15}_{-      4.38} $  &      1.63&&      1.12&\\
IC0724&    700.00& $^{+      0.00}_{-      0.00} $  &      0.02& $^{+      0.00}_{-      0.00} $  &     45.76&     18.62& $^{+      0.42}_{-      0.41} $  &      0.48&&     36.72&\\
IC1862&    700.00& $^{+      0.00}_{-      0.00} $  &      0.03& $^{+      0.00}_{-      0.00} $  &     29.56&      9.23& $^{+      5.34}_{-      5.97} $  &      1.07&&      0.76&\\
IC2531&    700.00& $^{+      0.00}_{-      0.00} $  &      0.03& $^{+      0.00}_{-      0.00} $  &     34.93&     15.53& $^{+      0.13}_{-      0.14} $  &      0.64&&   1602.52&\\
NGC0973&    685.65& $^{+      0.00}_{-      0.00} $  &      0.02& $^{+      0.00}_{-      0.00} $  &     44.35&     17.99& $^{+      1.47}_{-      1.60} $  &      0.70&&   3228.44&\\
NGC4705&    200.00& $^{+      0.00}_{-      0.00} $  &      0.09& $^{+      0.00}_{-      0.00} $  &     35.44&     11.81& $^{+      1.61}_{-      1.53} $  &      1.00&&      3.57&\\
NGC5529&    700.00& $^{+      0.00}_{-      0.00} $  &      0.05& $^{+      0.00}_{-      0.00} $  &     28.76&     16.33& $^{+      0.63}_{-      0.59} $  &      0.49&&    393.30&\\
NGC7184&    200.00& $^{+      0.00}_{-      0.00} $  &      0.33& $^{+      0.00}_{-      0.00} $  &     27.14&     26.02& $^{+      4.58}_{-      9.00} $  &      1.29&&      2.51&\\
PGC006966&    100.00& $^{+      0.00}_{-      0.00} $  &      0.91& $^{+      0.00}_{-      0.00} $  &     28.25&     32.58& $^{+      3.77}_{-      3.61} $  &      1.00&&      0.00&\\
UGC02885&    442.79& $^{+      0.00}_{-      0.00} $  &      0.09& $^{+      0.00}_{-      0.00} $  &     41.53&     39.68& $^{+     10.26}_{-      3.28} $  &      0.60&&     22.15&\\
UGC04277&     68.68& $^{+     17.40}_{-     43.25} $  &      1.16& $^{+      6.72}_{-      0.34} $  &     32.12&     30.65& $^{+     11.46}_{-      0.97} $  &      0.80&&   1148.61&\\
UGC05341&    100.00& $^{+      0.00}_{-      0.00} $  &      0.99& $^{+      0.00}_{-      0.00} $  &     35.28&     51.71& $^{+      0.55}_{-      1.61} $  &      0.37&&      0.00&\\
\hline
\end{tabular}
\end{center}
\end{table*}

\subsection{Notes on individual galaxies}\label{notes}
{\bf ESO030-009} is spiral galaxy for which no \HI data were available in literature. The disc and bulge I-band  mass-to-light ratios were fixed in a range 1.95 - 2.82 according to model $M/L$-colour relations and the $B-V$ colour inside of effective radius taken from Hyperleda data base and corrected for extinction in Galaxy. The resulting model of the rotation curve has maximum possible contribution of the disc. All halo density profiles give the infinite contours of  $1\sigma$ confidence levels on the $\chi^2$ map.\\
{\bf ESO184-051} is barred spiral giant galaxy without available  \HI radial density profile. For this case I firstly used I-band mass-to-light ratios in a range 1.79 - 2.46 according to observed effective $B-V$ colour index. However, this range leaded to the model of rotation curve with zero-contribution of the dark matter halo. Such model seems to be not very realistic, as far as baryonic contribution usually fails to reproduce the flat rotation curve outside of the optical radius (the region not covered by observations of this galaxy).  Thus I decided to decrease the values of possible mass-to-light ratios and had more reasonable contributions of dark and baryonic matter to the total mass of the galaxy without the infinite contours of  $1\sigma$ confidence levels on the $\chi^2$ map. \\
{\bf ESO240-011}  is spiral galaxy seen edge-on. For this system both optical and \HI kinematics are available together with the stellar kinematics. The rotation curve is very uncertain in the innermost part (see Fig. A1 in \citealt{Kregeletal2004}) thus I used the rotation velocity data for the radii $R>15$ arcsec.  The estimates of the stellar velocity dispersion at two disc radial scalelengths enabled  \cite{Zasovetal2011} to obtain the surface density of the stellar disc using the marginal gravitational stability criterion. This estimate corresponds to the disc I-band mass-to-light ratio 1.59, I took this value as the upper limit of disc M/L. The lower limit is 1.19 according to the observed colour index of the galaxy. The disc mass-to-light ratio estimated using gravitational stability criterion appears to be in good agreement with observed rotation curve. It indicates  the absence of strong dynamical overheating of the disc of ESO240-011. I show the model of the rotation curve of ESO240-011 and the $\chi^2$ map in Fig.  \ref{map_b}. One can see that despite the model of the rotation curve of ESO240-011 looks quite reasonable, the degeneracy of the central density and the radial scale of the halo does not allow for reliable estimate of the dark halo parameters.\\
{\bf ESO382-058} is giant spiral galaxy with a bar and a ring without available \HI radial distribution. The observed effective $B-V$ colour index corresponds to the range of $M/L_I$: 1.9 - 2.47. The resulting model is in satisfactory agreement with observations and $\chi^2$ shows finite contours of  $1\sigma$ confidence levels for all considered dark halo density profiles. \\
{\bf ESO460-031} is edge-on spiral galaxy without available \HI radial distribution. The observed effective $B-V$ colour index corresponds to the range of $M/L_I$: 1.68 - 2.22. The resulting model is in satisfactory agreement with observations, however $\chi^2$ demonstrates infinite contours of  $1\sigma$ confidence levels for all considered dark halo density profiles.\\
{\bf ESO471-002} is giant spiral galaxy without available \HI radial distribution. The observed effective $B-V$ colour index corresponds to the range of $M/L_I$: 1.8 - 2.46. The resulting model is in satisfactory agreement with observations however infinite contours of  $1\sigma$ confidence levels on the $\chi^2$ maps for all considered halo density profiles do not allow me to estimate reliably the parameters of the dark halo.\\
{\bf ESO563-021} is spiral giant galaxy  also presented in the sample of \cite{Courtois2015}. 
 Its stellar mass-to-light ratio was fixed in the range:  1.46 - 1.74 according to its $B-V$ colour inside of effective radius. The model of the rotation curve is in satisfactory agreement with observational data and the contours of  $1\sigma$ confidence levels are finite for all considered dark halo density profiles except NFW.\\
{\bf IC0724} is Sa-type galaxy with no available \HI radial distribution. The radial profile of stellar velocity dispersion obtained by \cite{Corsinietal1999} allows to estimate independently the upper limit of the disc surface density using marginal gravitational stability criterion (see \citealt{Zasovetal2011}). The disc K-band mass-to-light ratio related to this estimate lies in the range 0.48 - 1.22 (corresponding to the errors of the observational data). These values are very close to that following from the stellar population synthesis models indicating that the disc of IC0724 is not dynamically overheated. The models of \cite{bdj} give $M/L_K$ of stellar population that lies in the range 0.4-0.8. \cite{McGaughSchombert2014} gives: $M/L_K=0.6$ very slightly varying with the colour index. For the disc of IC0724 I utilized the range of mass-to-light ratios determined by marginal gravitational stability criterion and for the bulge I decided to use the range of $M/L_K$: 0.6-0.8. The resulting model of the rotation curve is a model with maximal contribution of the baryonic matter, the contours of  $1\sigma$ confidence levels are infinite for all considered profiles of dark halo.\\
{\bf IC1862}  is edge-on spiral galaxy without available information on the radial distribution of \HI surface density. The mass-to-light ratio in the I-band according to the effective $B-V$ colour lies in the range: 1.88 - 2.64. However these values lead to the overprediction of the rotational velocity amplitude so I had to decrease the lower limit of mass-to-light ratio by 2 times in order to get reasonable model of the rotation curve. The $\chi^2$ map has infinite contours of  $1\sigma$ confidence levels for all considered density profiles of dark halo.\\
{\bf IC2531} is another edge-on late type spiral galaxy. \cite{Zasovetal2011} estimated the upper limit of the stellar disc surface density of this galaxy using the marginal gravitational stability criterion. After correction for the  adopted distance it leads to the following value of V-band disc mass-to-light ratio: $M/L_V = 0.64$. It appears to be lower than follows from the observed colour index: 1.06-1.43 giving evidences in favor of the absence of significant major merging events, which could heat dynamically the inner parts of disc. For the disc of IC2531 I fixed mass-to-light ratios in the interval: 0.64-1.43, for the bulge I utilized the values coming from the colour index.
The resulting model is in a good agreement with the observed rotation curve. The $\chi^2$ map has finite $1\sigma$ contours for all considered dark halo density distributions except NFW.\\
{\bf NGC0973} is giant spiral galaxy seen edge-on. The photometrically determined contribution of stars to the rotation curve significantly overpredicts the rotation velocity amplitude, thus for this galaxy I decided to use mass-to-light ratios of disc and bulge as free parameters. The resulting best-fitting model is in good agreement with observations but the $1\sigma$ contours are infinite for the $\chi^2$ maps plotted for all considered dark halo density distributions. \\
{\bf NGC4705} is barred spiral galaxy with a ring and no available data on the \HI surface density distribution. The effective $B-V$ colour corresponds to the I-band mass-to-light ratio range of 1.8 - 2.46. However these values for the disc lead to the overprediction of the observed rotation curve. Thus, I had to decrease the lower limit of the disc mass-to-light ratio to 1. The resulting model gave the satisfactory fit to the observed rotation curve and the $1\sigma$ contours  are finite for piso and Burkert  dark halo density distributions.\\
{\bf NGC5529}  is barred spiral galaxy. For NGC5529 I found both optical and \HI kinematical data. The $H\alpha$ kinematical data from \cite{Kregeletal2004} is in the contradiction with the \HI data from \cite{Allaertetal2015} in the inner part of the galaxy, making the rotation curve uncertain in the inner region, so I decided to remove the innermost points (R<4 kpc) from consideration. The stellar velocity dispersion data enabled \cite{Zasovetal2011} to estimate the upper limit of the disc surface density related to its marginal gravitational stability at 2 disc radial scalelengths. This estimate corresponds to the disc V-band mass-to-light ratio $M/L_V=0.54$ which is lower than follows from the observed colour index, according to the equations from \cite{McGaughSchombert2014}: 1.31-1.66. It can indicate the absence of dynamical overheating of the disc of NGC5529. However, the stellar mass-to-light ratios obtained from both gravitational stability criterion and observed colour index for the disc and using only photometry for the bulge together with the surface photometry from \cite{Xilouris1999} give overpredicted value of the rotational velocity. Thus I had to decrease the lower limit of mass-to-light ratio by 10 percent for the disc and by 1.5 times for the bulge. After this reduction I got the reasonable fit of the rotation curve and the finite $1\sigma$ contours   on the $\chi^2$ maps for piso and Burkert dark haloes.\\
{\bf NGC7184 } is barred giant spiral galaxy with a ring and no available \HI data. The observed total colour index corresponds to the I-band mass-to-light ratio in the range 1.29 - 1.43. The resulting fit of the rotation curve is in satisfactory agreement with observed data, however the $1\sigma$ contours   on the $\chi^2$ maps  are infinite for all considered density profiles of dark halo.\\
{\bf PGC006966} is Sc type edge-on giant galaxy without available radial profile of \HI  surface density. The effective $B-V$ colour refers to the I-band mass-to-light ratio range: 1.8 - 2.46. However, the rotation velocity increases very slowly in the central part leaving almost no room for the bulge. So I had to reduce the lower limit of mass-to-light ratio of the disc to 1 and set the bulge contribution to zero which could be reasonable due to the low contribution of the bulge to the total luminosity. The resulting fit of the rotation curve is in satisfactory agreement with observations but not for NFW density profile. The $1\sigma$ contour for the $\chi^2$ map plotted for NFW halo is also infinite.\\ 
{\bf UGC02885}  is giant Sc galaxy. It is also present in the sample of \cite{Courtois2015}. The parameters of piso dark halo agree with that from \cite{deBlok1997} within error bars. The $3.6\mu \mathrm{m}$   mass-to-light ratios of disc and bulge were taken to lie between 0.45 \citep[obtained using Tully-Fisher relation in a good agreement with the models of stellar population by][]{McGaugh2015} and 0.6 \citep{Meidtetal2014} solar units. The model rotation curve is in reasonable agreement with observations. The $1\sigma$ contours for the $\chi^2$ maps are finite for piso and Burkert haloes.\\
{\bf UGC04277}  is Sc type edge-on giant galaxy.  The K band disc mass-to-light ratio is set within the range: 0.37-0.8 solar units (according to \citealt{bdj}). Due to the lack of photometrical data the  budge central density, effective radius and Sersic index are considered as free parameters\footnote{The resulting parameters for the bulge: n=4, $R_e=2.7$ kpc.}. The model rotation curves and  $\chi^2$ maps for this galaxy are demonstrated in Fig. \ref{map_a}. One can see that the model rotation curve is in good agreement with observed one and $1\sigma$ confidence limit contours for the $\chi^2$ maps are finite for all dark halo profiles.\\ 
{\bf UGC05341} is Sc type edge-on giant galaxy without available \HI surface density radial distribution. The K band disc mass-to-light ratio is set within the range: 0.37-0.8 solar units (according to \citealt{bdj}). The slowly rising rotation curve in the inner part of the galaxy does not leave room for a bulge, so I neglected its contribution to the total mass. The resulting model rotation curve is in reasonable agreement with observations for piso and Burkert dark halo. The $1\sigma$ contours for the $\chi^2$ maps are also finite for these halo profiles.\\

One can see that for 6 galaxies from the sample I had to decrease the lower limit of the disc mass-to-light ratio due the fact that photometricaly determined disc surface density  leaves no room for dark matter halo or even corresponds to the model rotation velocity amplitude which is higher than the observed one. Unfortunately this situation can occur even for well studied nearby galaxies (see e.g. the discussion of this problem in \citealt{saburovaetal2016}). I compared these 6 systems with the contradiction between photometrical and dynamical estimates of the disc surface density and found out that 3 of them are classified in Hyperleda database as barred galaxies (in total I have only 5 objects with such classification in my sample and 3 of them are among the galaxies with super-maximal discs). The remaining 3 super-maximal disc galaxies are seen edge-on so their classification could be difficult, but I inspected their images and found boxy-shaped bulges in two of them which can speak in favour that they are also barred (see e.g. \citealt{ChungBureau2004}). Thus the most probable reason of the discrepancy between dynamical and photometrical disc mass estimates is the presence of the bar which can lead to lower rotation velocity (see e.g. \citealt{Saburovaetal2017}).
Another item that is related to this problem is to what extent the model dark halo parameters are sensible to the disc mass. Partially this problem is discussed in \cite{saburovaetal2016} where we considered both best-fitting modeling of the rotation curve when the surface densities of disc and bulge were considered as free parameters and the photometrical approach when the baryonic densities were fixed in the narrow ranges. According to \cite{saburovaetal2016} the dark halo mass is the most stable parameter which is less affected by the choice of the model.  Since some of my conclusions are based on the dark halo parameters I also tested how they change with the decrease of the disc surface density on the example of NGC4705 for Burkert profile. As it is expected when I decrease the disc surface density the halo parameters are also changing in order to eliminate the increasing difference between the model and observed rotation curve in the region of the high contribution of the disc. The dark halo radial scale becomes lower and its central density increases, the mass of the dark halo also rises. In agreement with \cite{saburovaetal2016} the dark halo mass is the most stable parameter. It changes by 7 per cent when the disc density decreases by 10 per cent. Even when I decrease the disc density by 55 per cent it increases by roughly 30 per cent. The radial scale of the halo is less stable -- it     changes up to 20 per cent when the disc density is lowered by 10 per cent. If I decrease the disc density by 55 per cent the dark halo radial scale changes by roughly 50 per cent. The most unstable parameter is the central density of the halo which changes by almost 4 times in response to the variation of the disc density by 55 per cent. However even the most extremal changes of the parameters of dark matter halo due to the variation of the disc density lie in the range of the dispersion of points on the diagrams (see below) and the uncertainty of the parameters of dark halo is borne in mind.

\section{Discussion}\label{Discussion}

As it is evident from the details given above, the modeling of rotation curves of giant discy galaxies using NFW dark halo often leads to the infinite $1\sigma$ confidence limit contours for the $\chi^2$ maps. It means that it is not possible to get reliable estimate of dark halo central density and radial scale for NFW profile based on the available data. Thus, I decided to eliminate the NFW parameters from further consideration.  I can not exclude the possibility that NFW profile fails more often in comparison to other profiles to reproduce the rotation curves of giant galaxies, due to e.g. the observed slow rise of the velocity, but more accurate data are needed especially in the central regions to confirm or disprove this possibility. This will be a part of my future work.   

\cite{Kasparova2014} proposed that the giant LSB galaxy Malin2 could have such large disc due to the peculiar properties of its host dark matter halo, which has high radial scalelength. I decided to find out if the dark matter haloes of giant discy galaxies of my sample differ from that of the systems with more moderate sizes. One should remember, however, that the dark halo central density and radial scale are known to be very uncertain due to the degeneracy between these parameters (see \citealt{saburovaetal2016}). I partly solve this problem by fixation of the baryonic densities in the narrow ranges and by considering the $\chi^2$ map for the dark halo parameters. However,  these parameters still should be considered with great caution.  \cite{saburovaetal2016} showed that the dark halo mass within optical border is much more reliable parameter, so I also discuss the dark halo masses of giant discy galaxies.

In Figs. \ref{rho} - \ref{mhalo} I plot the dark halo central densities, radial scales and masses within optical borders against the disc sizes. Squares show the compilation of literature data on the dark halo parameters from \cite{Saburova2014}, which also contains three giant LSB galaxies Malin 1,  Malin 2 and NGC7589 for piso dark halo. Diamonds demonstrate the data from the current paper. For the current paper I plotted the  four disc scalelengths as disc radius. For data from \cite{Saburova2014} I considered the radius of 25-th B-band isophote as the size of the disc for all galaxies except LSBs for which I also took four disc scalelengths. Left panels correspond to piso halo, right ones are for Burkert halo. The error bars for the sample of giant galaxies are associated with the range covered by $1\sigma$ confidence limit. The lines show linear fit to the data. I did not use giant galaxies with infinite $1\sigma$ countours in the fitting, they are also absent in Figs. \ref{rho} - \ref{mhalo}.

 As one can see, there are clear trends of the parameters of Burkert halo with the optical radius. For piso halo the points have higher spread in comparison to Burkert halo. Partly it can be explained by the fact that for the Burkert halo one can get reliable estimates of the density and radial scale slightly more often than for piso and NFW halo (see \citealt{saburovaetal2016}).   Bearing in mind the uncertainty of the estimates of density and scale of dark haloes based on the rotation curve decomposition one can still see that the central halo density decreases with the optical radius, and the scalelenght increases with it for Burkert profile. I found the following correlations between the parameters: $\log(\rho_0)=-1.14-0.72\log(R_{opt})$  (correlation coefficient 0.60); $\log(R_s)=0.01+0.86\log(R_{opt})$ (correlation coefficient 0.83). The relation between the sizes of baryonic and dark matter structures could possibly be explained by the connection of the specific angular momentum of the galaxies and the host dark haloes (see e.g. \citealt{Moetal1998}). \cite{Kravtsov2013} found approximately linear relation between the radius containing half of stellar mass and the virial radius estimated  using abundance matching ansatz. Thus the almost linear relation between the sizes that I obtained are expected in the models of the formation of the galaxies. Considering the correlation with the central density it may be  due to the connection between central density and radial scale of the halo (see, e.g. \citealt{Kormendy2004}, \citealt{Burkert2015} and discussion in \citealt{Saburova2014} and \citealt{saburovaetal2016}).

For the dark halo mass the correlation is more significant and firmly determined in comparison to that with the radial scale and density for both Burkert and piso haloes. I found the following correlation for Burkert halo: $\log(M_h(R_{opt})=8.25+2.13\log(R_{opt})$ (correlation coefficient 0.97) for the piso halo the result is very similar: $\log(M_h(R_{opt})=8.25+2.12\log(R_{opt})$ (correlation coefficient 0.89). Partially this correlation can be explained by the connection between the halo and stellar mass which could follow e.g. from the presence of Tully-Fisher relation (\citealt{Roberts1969}; \citealt{TullyFisher1977}). The relation between the virial halo mass and the disc mass for a sample of discy galaxies was demonstrated e.g. in \cite{Zasovetal2017}. As far as stellar mass is connected to the size of the disc it is natural to expect that the latter is also related to the halo mass. However, the index of the power of the relation I got in current paper is the same (within the uncertainty) as was found for the fixed stellar mass by \cite{Charltonetal2017}, who considered a relation between the effective radius and the dark halo mass estimated  using galaxy-galaxy lensing.  Thus, the correlation between the size of the disc and the halo mass could be not fully due to the connection between the masses.

 Another important thing is  that giant discy galaxies do not deviate significantly from the trends found for galaxies with more moderate sizes. In particular the giant spiral galaxies of my sample do not tend to have significantly lower halo masses compared to the general trend which could be expected in the formation scenario from the
inside out by accretion of cold gas proposed by \cite{Ogleetal2016}. \footnote{One should admit, however, that spiral giant galaxies of my sample have slightly lower luminosities than superluminous spirals so they could have different origin.}  It can indicate that the relations could be due to the same physical processes. Thus, e.g. the catastrophic scenarios of the formation of such galaxies can be ruled out since there is no need to introduce catastrophic scenario to explain the parameters of moderate size spirals which lie on the same trend. Another indication of absence of major merger events comes from the marginal gravitational stability of the several giant discs of my sample with available data on the stellar velocity dispersion.

 The found relations (especially the ones between  the radial scale and the central density with optical size) could speak in favour of the scenario proposed by \citet{Kasparova2014} in  which the giant size of the disc is explained by the properties of dark matter halo, namely the high radial scale of the halo. Another important factor can be the environment in the stage of the formation of giant galaxies -- one needs a lot of gas to form the massive HSB discs. This gas could be accreted from the largescale filaments.

It should be noted, however, that giant galaxies from my sample differ from giant LSBs like Malin2 -- they have considerably smaller sizes and higher surface brightness. At the moment, there is a lack of the estimates of dark halo parameters of giant LSBs to understand if they differ significantly from the giant galaxies of my sample. At the moment it seems that the parameters of giant LSBs lie on the same trend as giant HSBs, except the halo mass which seems to be slightly lower than the relation.   High quality data on the rotation velocities in the central parts of giant LSBs will allow to obtain more reliable estimates of dark halo parameters and to clarify this question.

\begin{figure*}
\hspace{-1.5cm}
\includegraphics[width=0.35\textwidth]{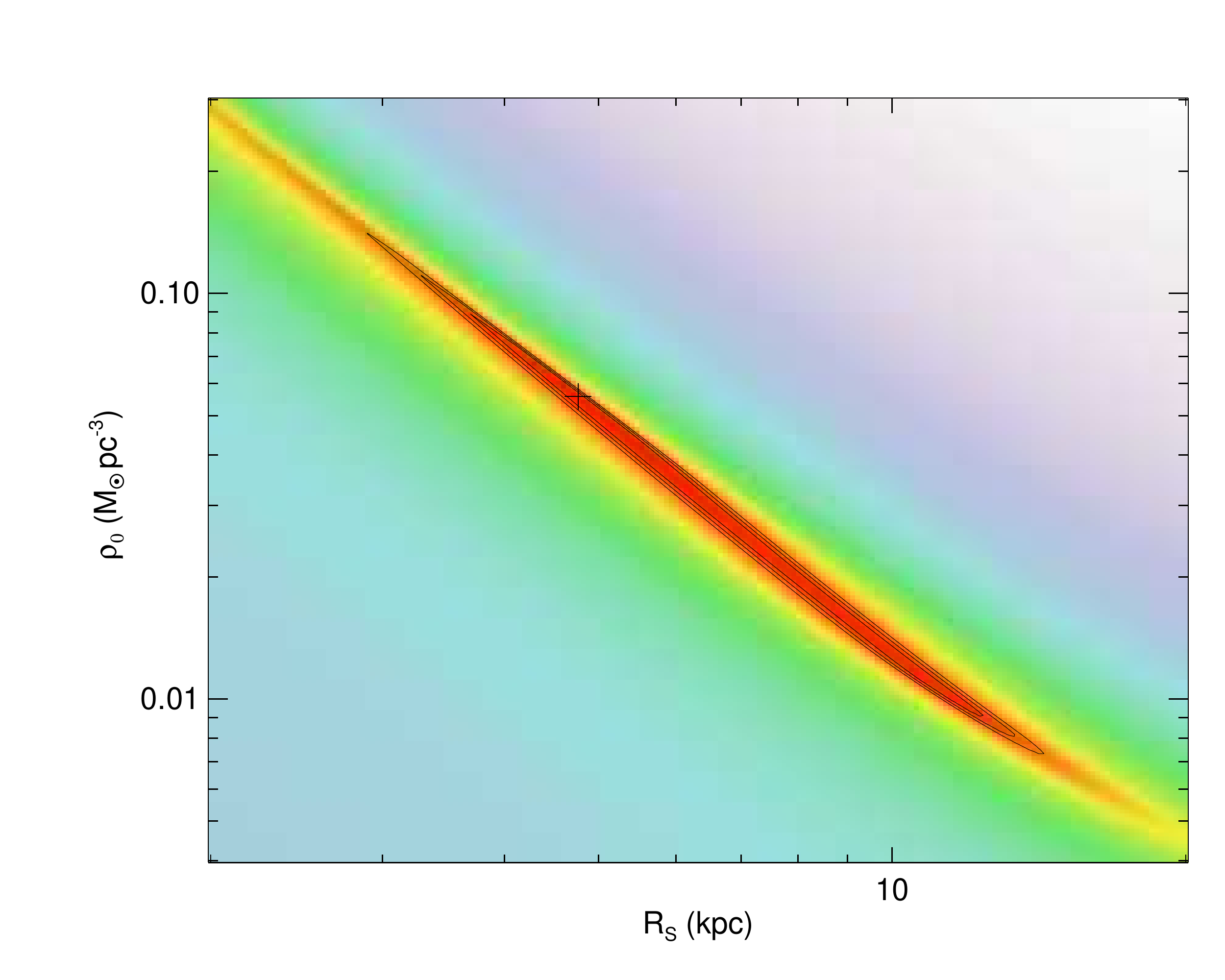}
\includegraphics[width=0.35\textwidth]{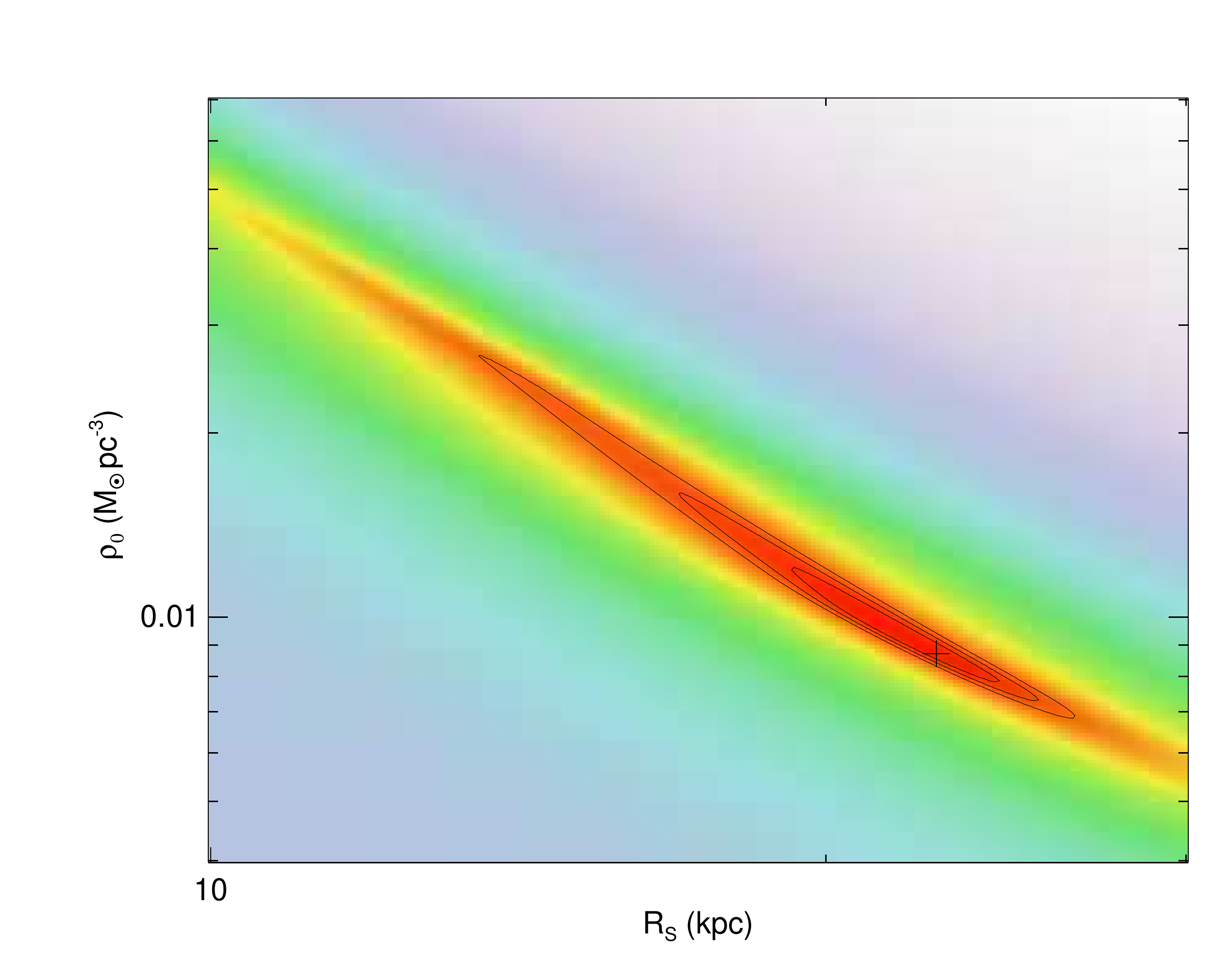}
\includegraphics[width=0.35\textwidth]{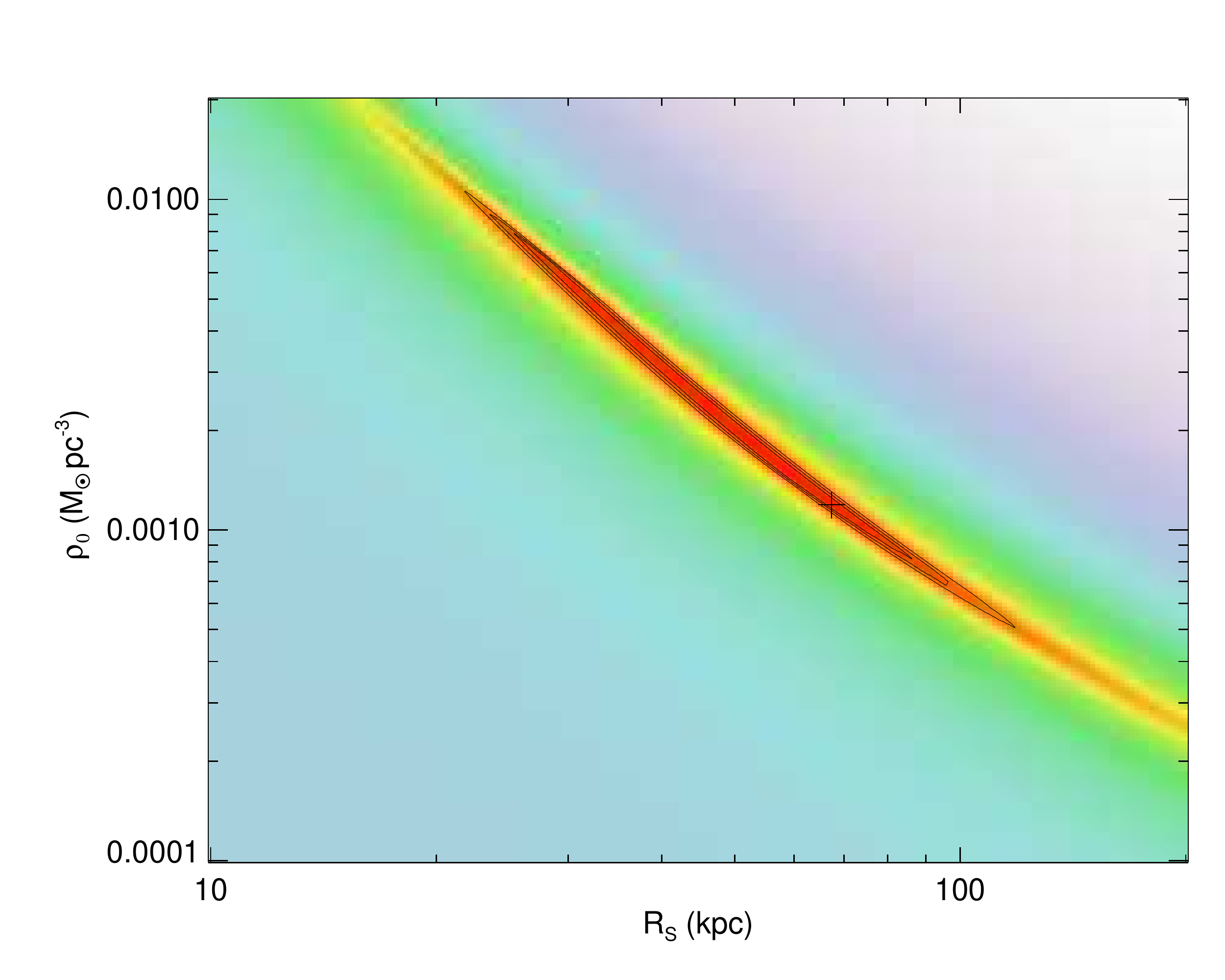}
\hspace{-18.5cm}
\includegraphics[height=0.35\textwidth,angle=-90]{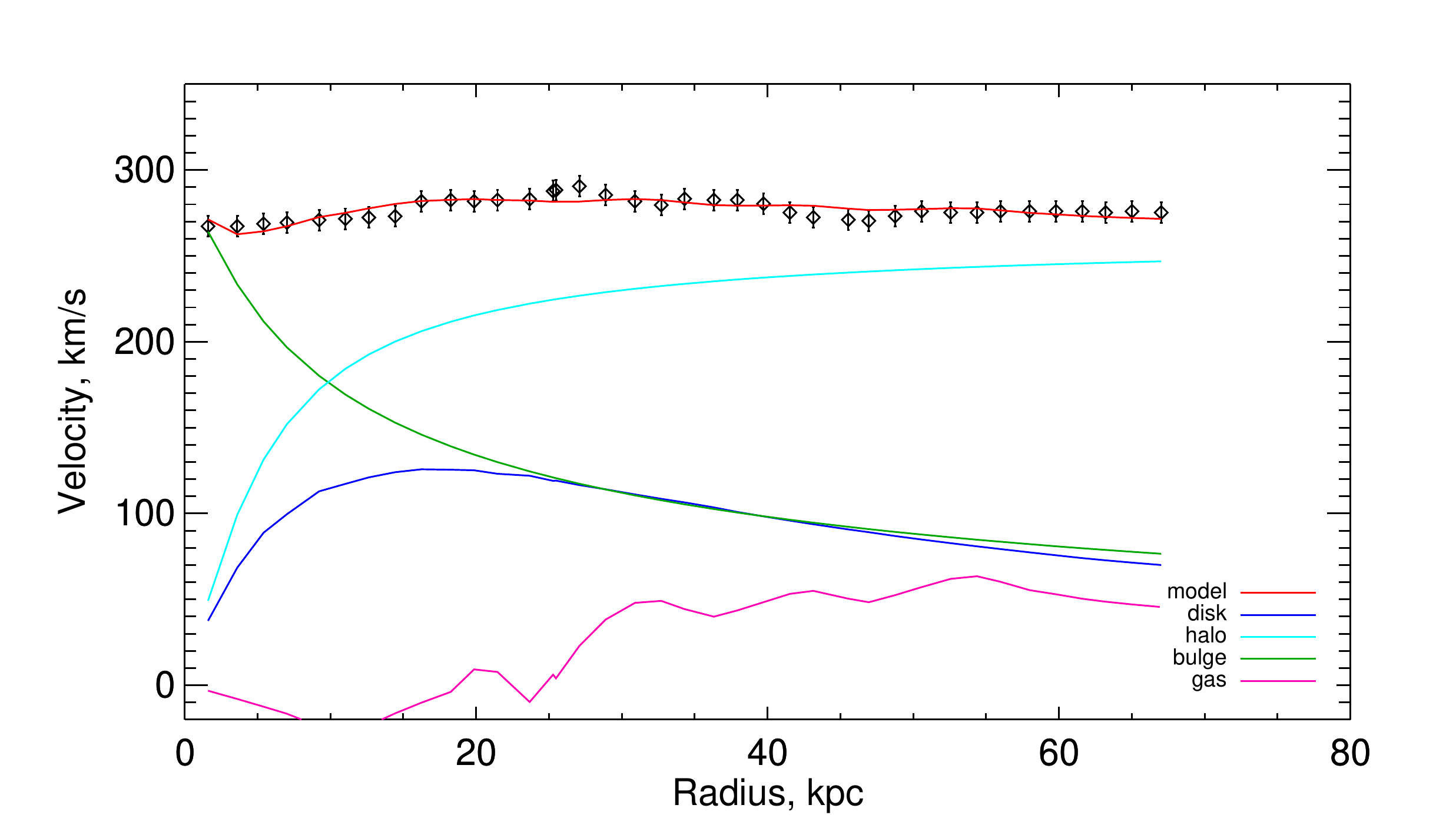}
\includegraphics[height=0.35\textwidth,angle=-90]{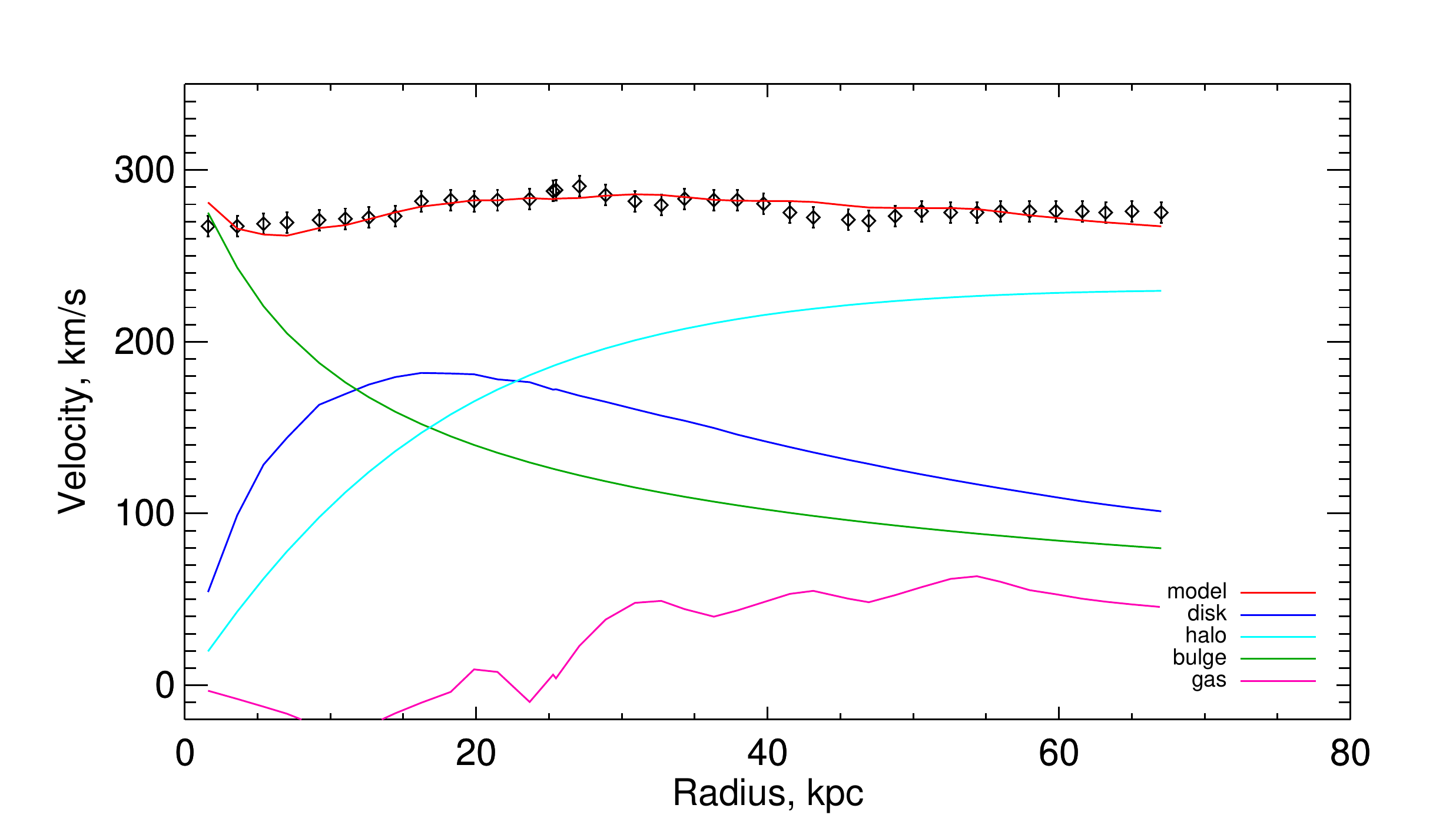}
\includegraphics[height=0.35\textwidth,angle=-90]{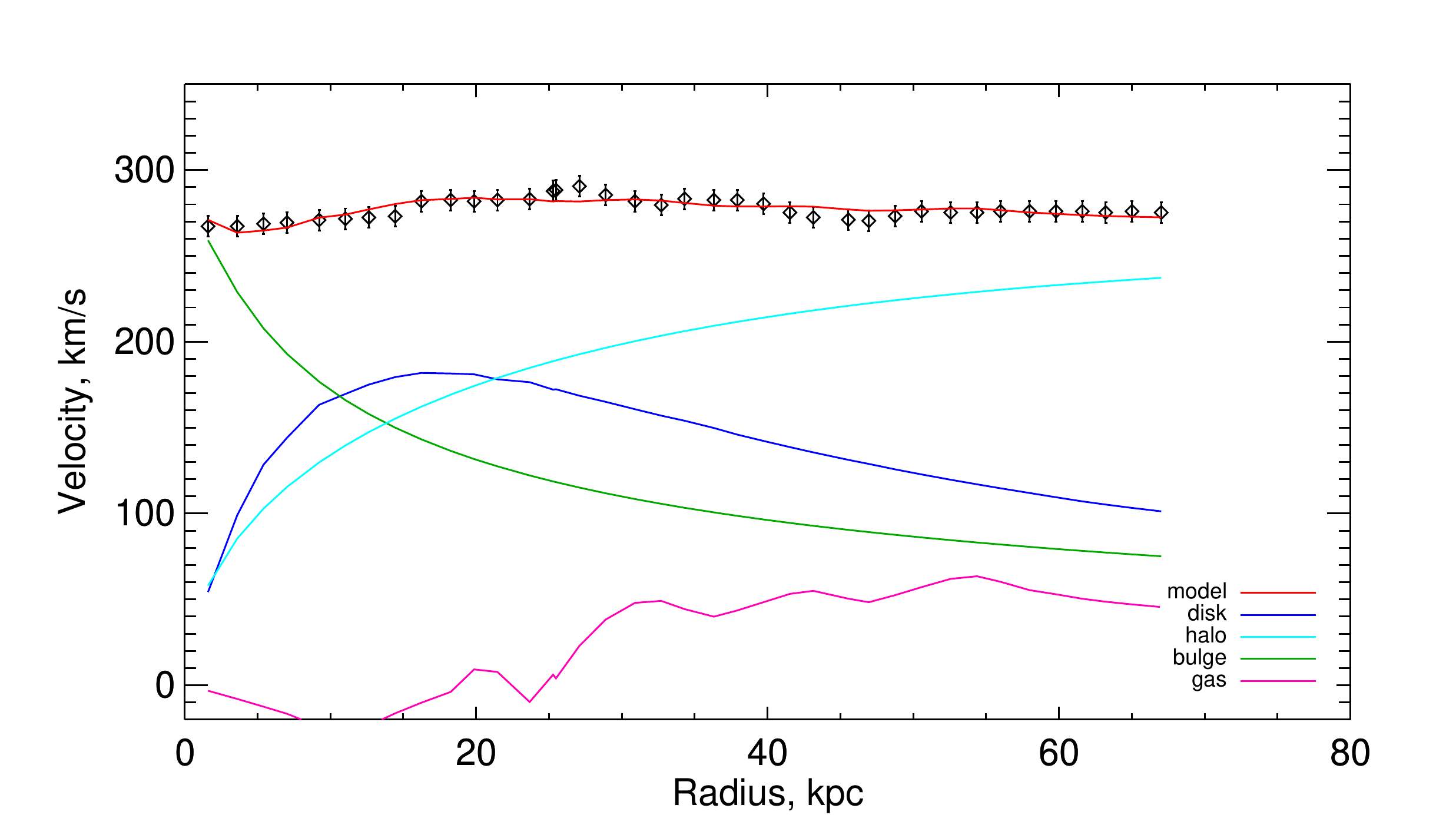}
\vspace{1cm}
\caption{ The $\chi^2$ maps and rotation curve models for UGC4277. Left panels correspond to the models with the piso profiles of the dark halo, centre panels~--- to the Burkert  profiles, right panels~--- to the NFW profiles.
The colour on the maps denotes the $\chi^2$ value, the redder the colour, the lower the $\chi^2$ and the better is the fit. 
The black contours refer to $1\sigma$, $2\sigma$ and $3\sigma$ confidence limits. The cross shows the position of the $\chi^2$ minimum. 
 The lower row  gives the best-fitting decomposition
corresponding to the $\chi^2$  minimum, where the black symbols  mark the observed rotation curve, thick red line~--- the total model, cyan  line~---  dark halo, blue line~--- stellar disc, magenta line~---  gas disc, green line~---  bulge.
}
\label{map_a} 
\end{figure*}
\begin{figure*}
\hspace{-1.5cm}
\includegraphics[width=0.35\textwidth]{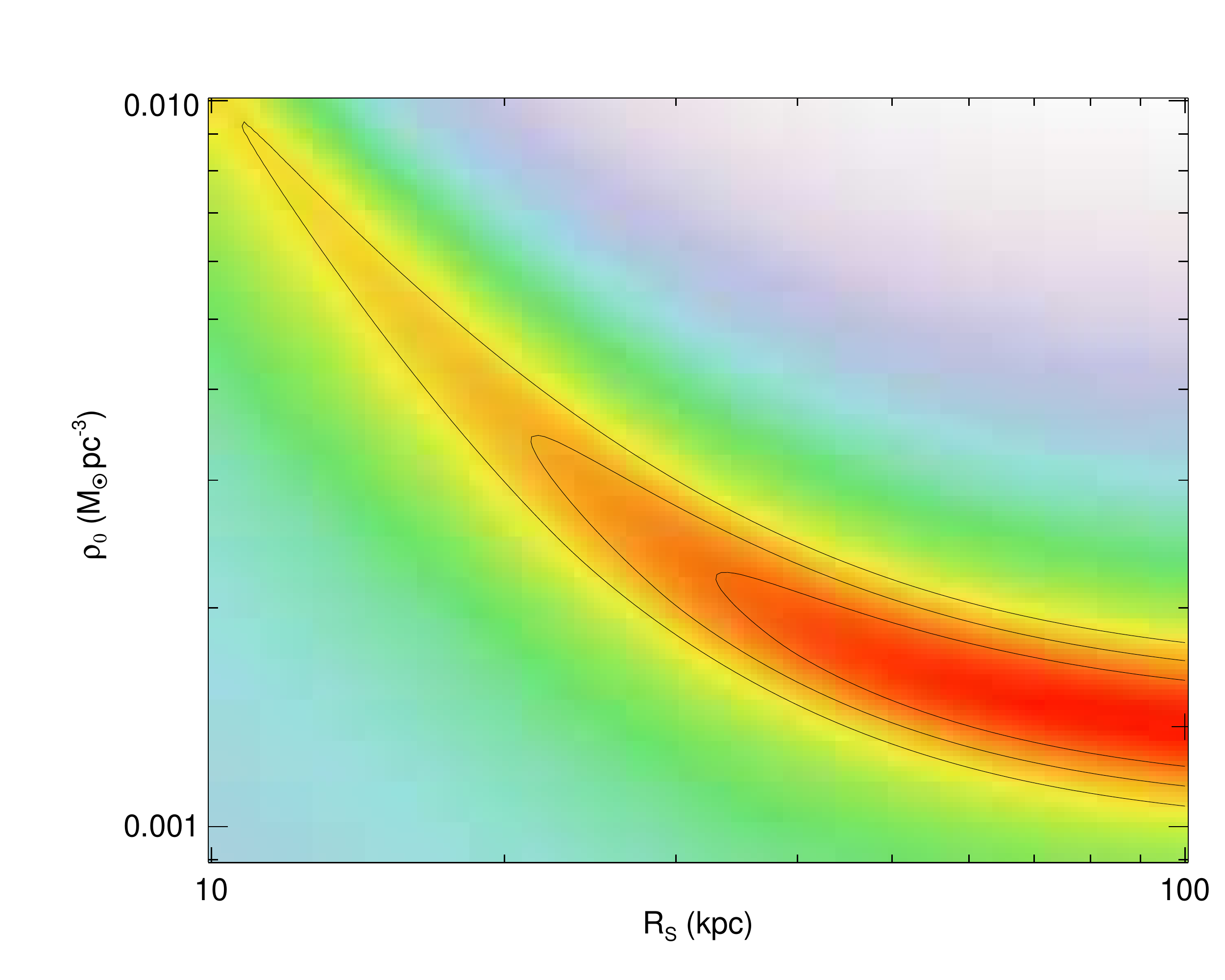}
\includegraphics[width=0.35\textwidth]{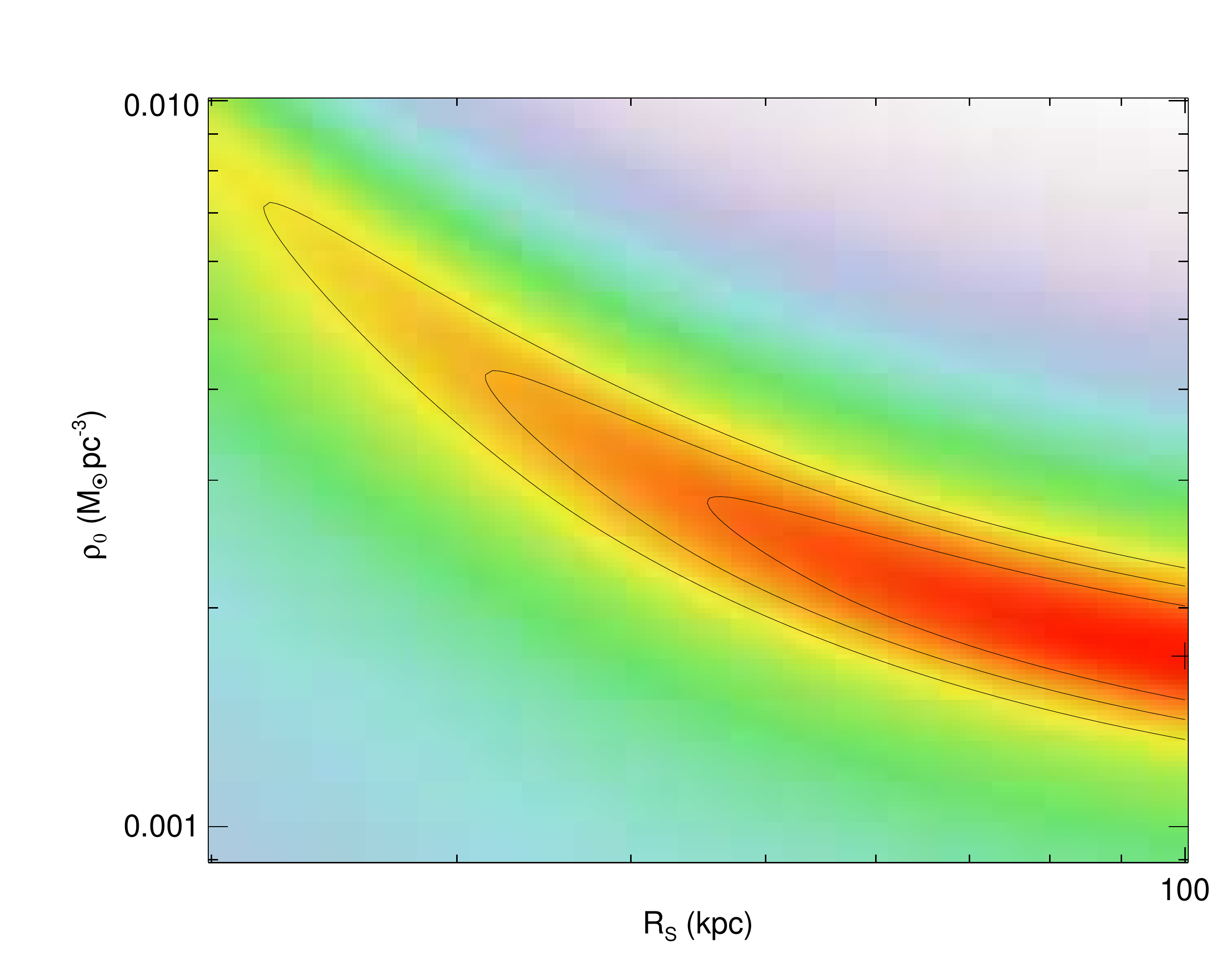}
\includegraphics[width=0.35\textwidth]{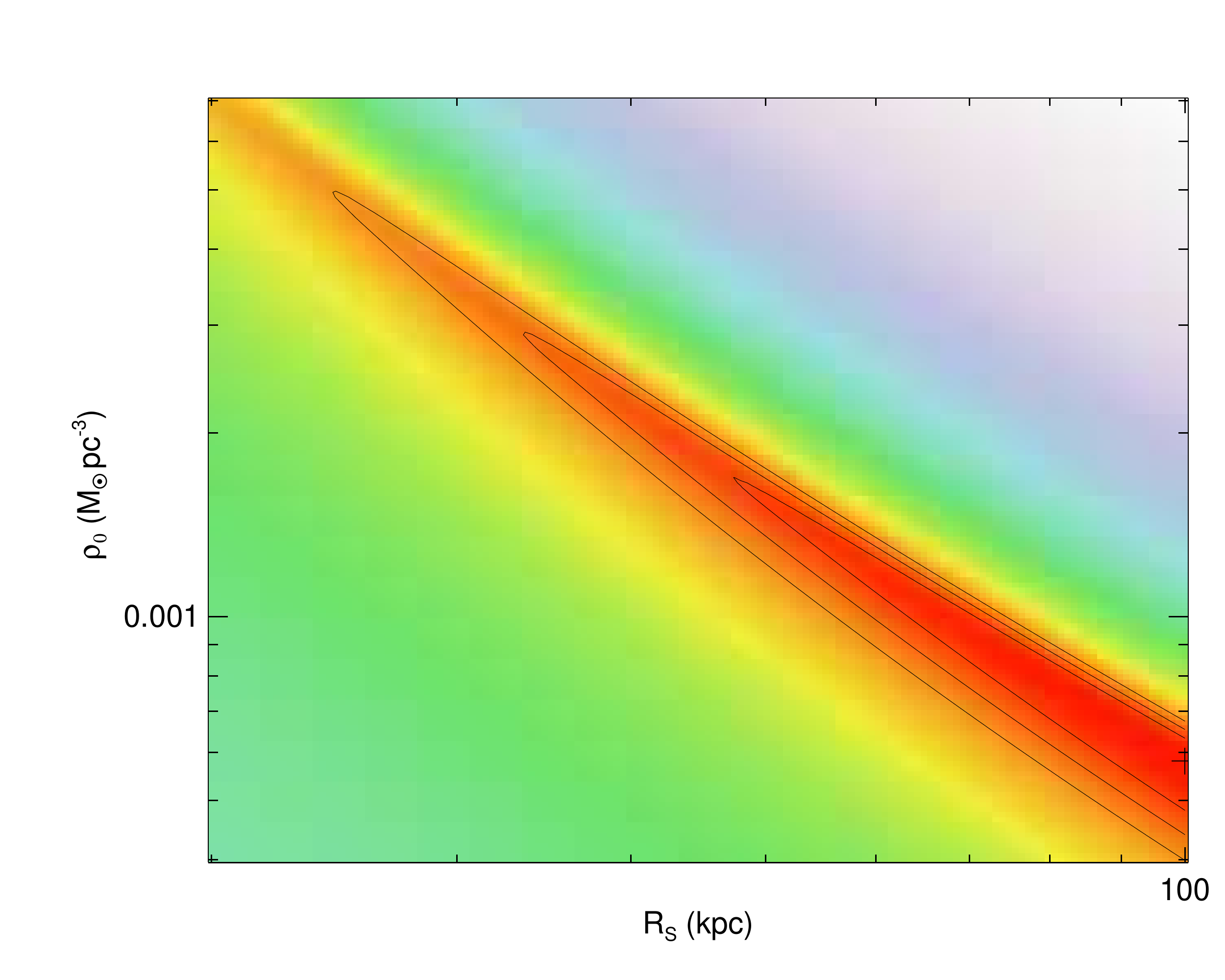}
\hspace{-18.5cm}
\includegraphics[height=0.35\textwidth,angle=-90]{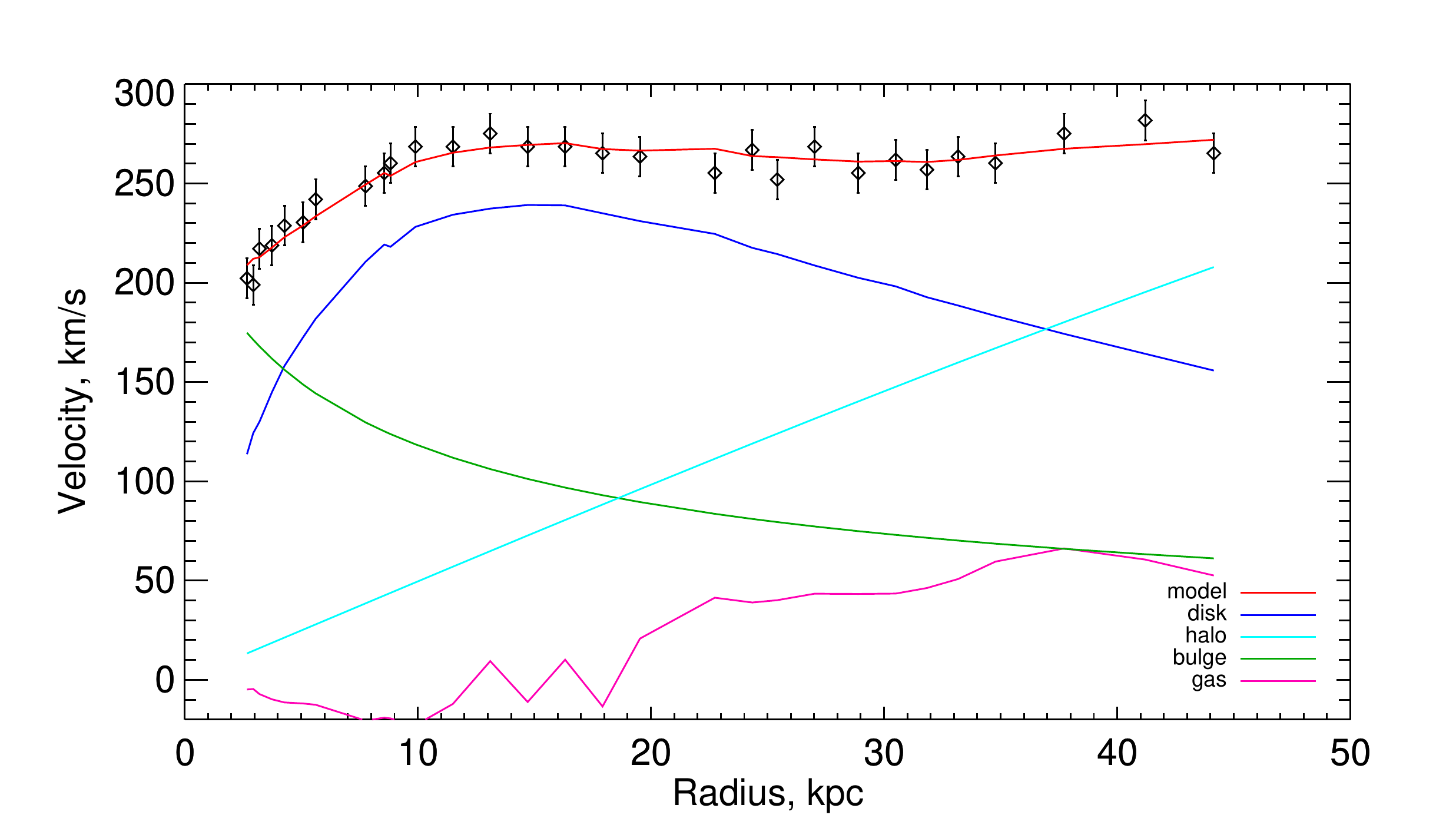}
\includegraphics[height=0.35\textwidth,angle=-90]{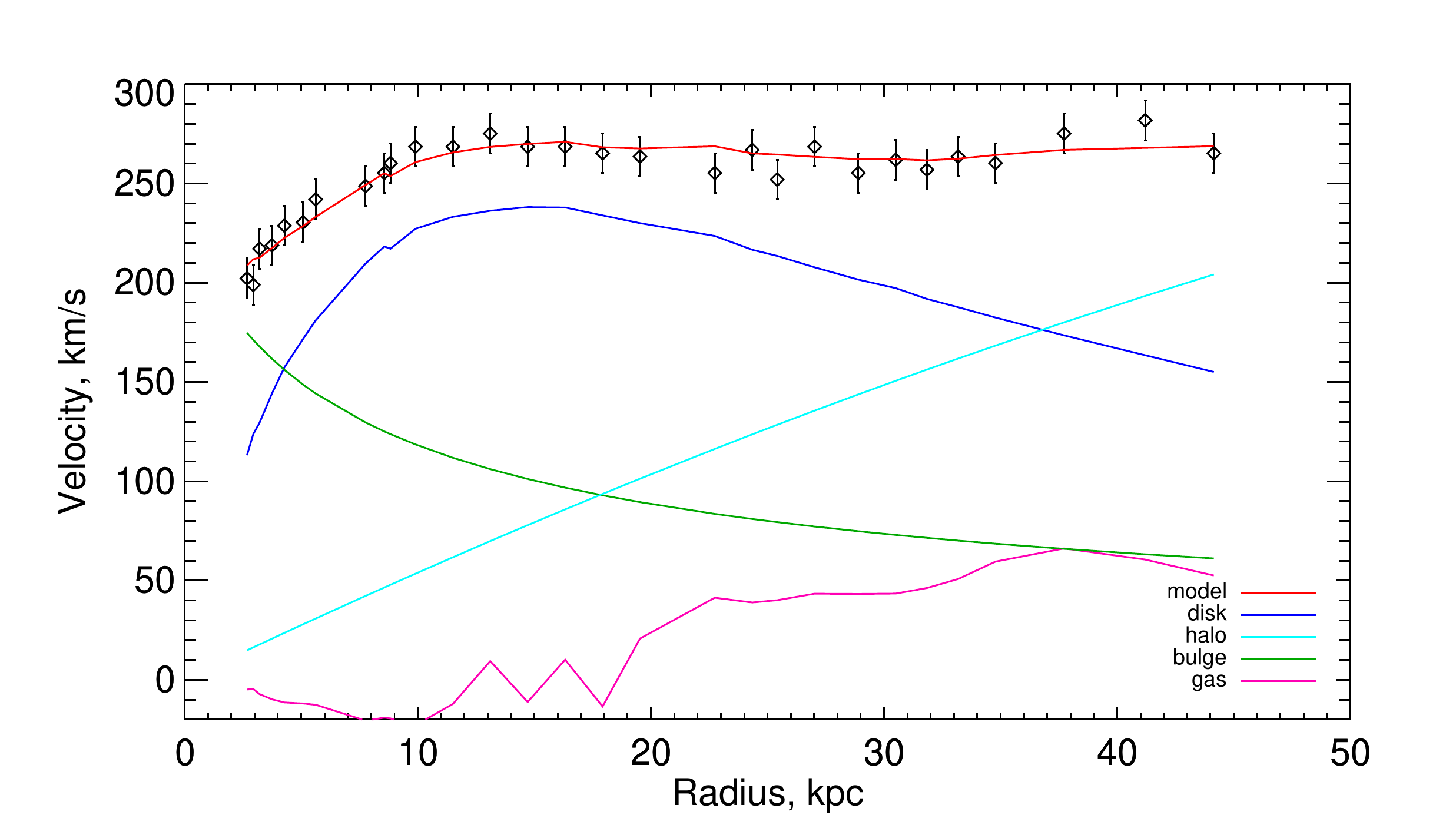}
\includegraphics[height=0.35\textwidth,angle=-90]{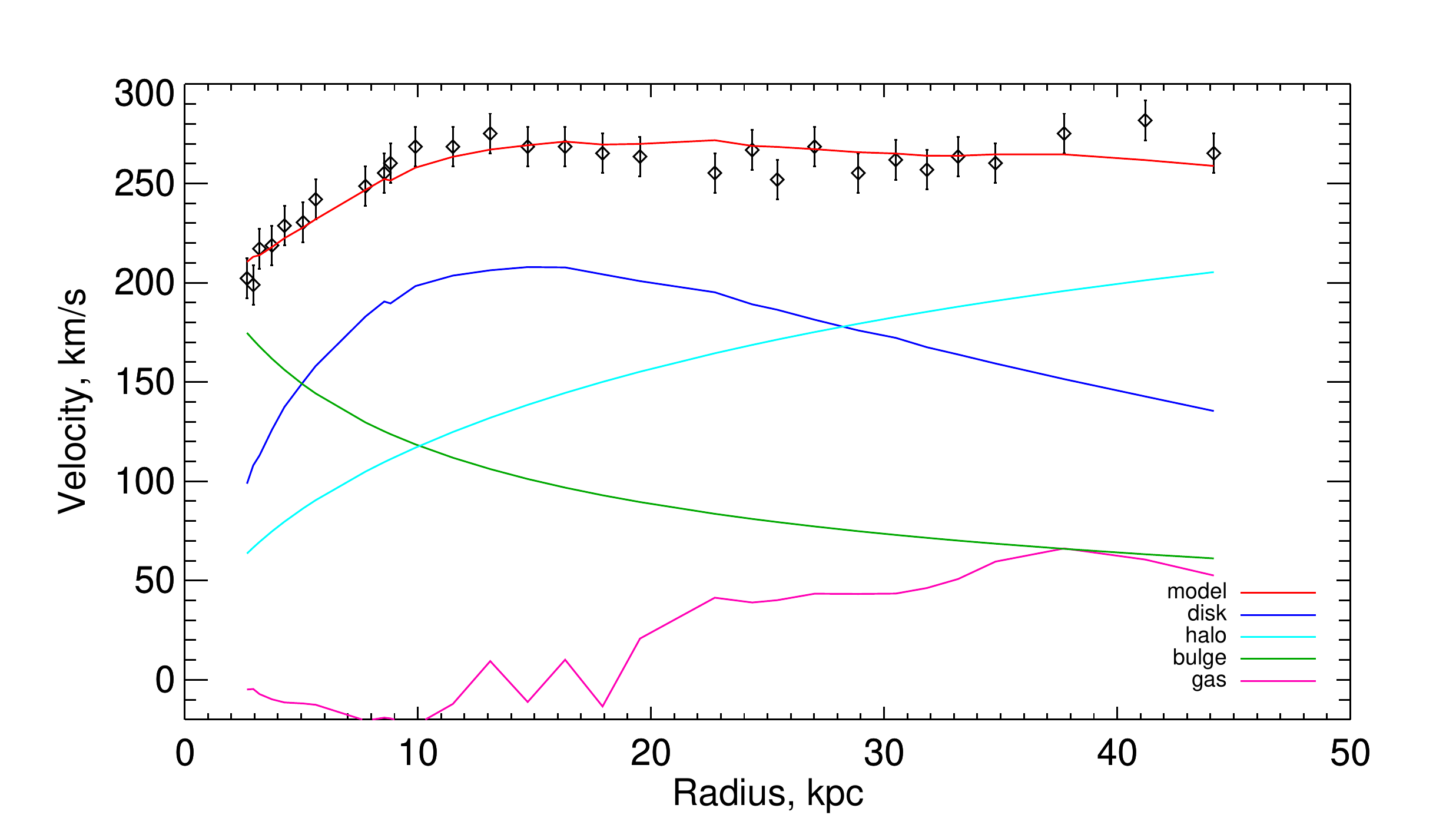}
\vspace{1cm}
\caption{ The $\chi^2$ maps and rotation curve models for ESO240-11. Left panels correspond to the models with the piso profiles of the dark halo, centre panels~--- to the Burkert  profiles, right panels~--- to the NFW profiles.
The colour on the maps denotes the $\chi^2$ value, the redder the colour, the lower the $\chi^2$ and the better is the fit. 
The black contours refer to $1\sigma$, $2\sigma$ and $3\sigma$ confidence limits. The cross shows the position of the $\chi^2$ minimum. 
 The lower rowgives the best-fitting decomposition
corresponding to the $\chi^2$  minimum, where the black symbols  mark the observed rotation curve, thick red line~--- the total model, cyan  line~---  dark halo, blue line~--- stellar disc, magenta line~---  gas disc, green line~---  bulge.
}
\label{map_b} 
\end{figure*}

\begin{figure*}
\includegraphics[width=1\textwidth]{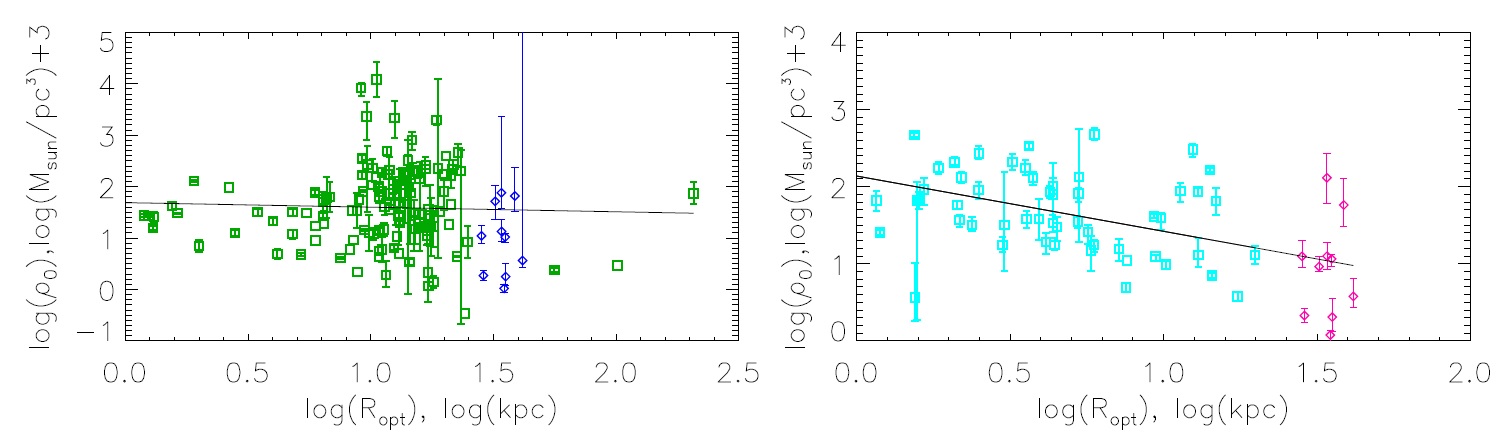}
\caption{ The comparison of central density of the halo with the optical radius for the current sample of giant spirals (diamonds) and the sample  from Saburova, Del Popolo (2014) (squares). The left hand panel corresponds to piso dark halo, the right one -- to Burkert halo. The error bars for the sample of giant galaxies are associated with the range covered by $1\sigma$ confidence limit. The line shows the linear fit to the data.}
\label{rho} 
\end{figure*}
\begin{figure*}
\includegraphics[width=1\textwidth]{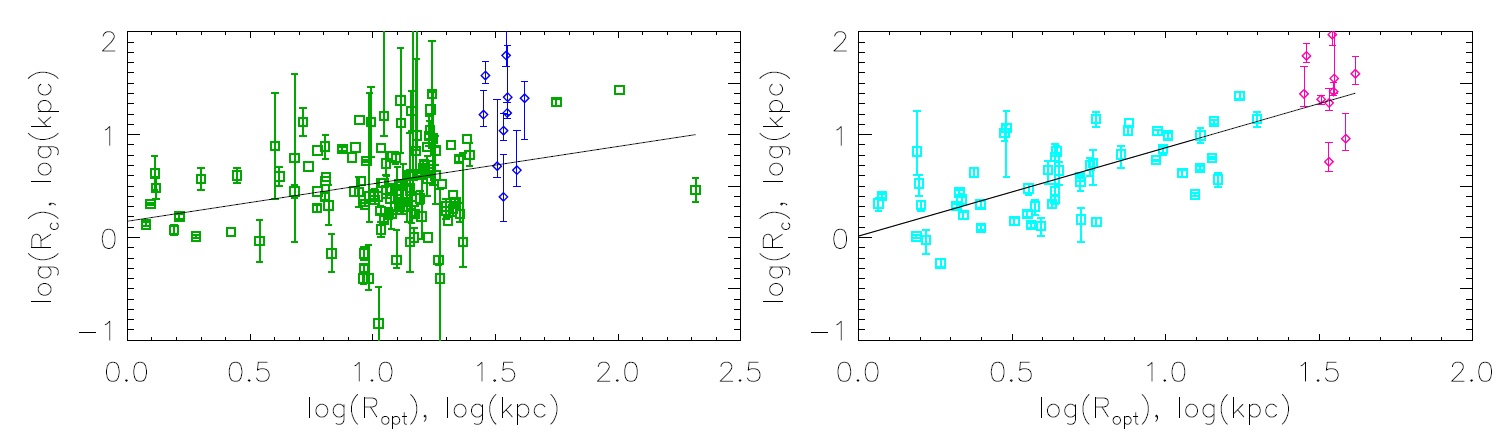}
\caption{ The comparison of radial scalelength of the halo with the optical radius for the current sample of giant spirals (diamonds) and the sample  from Saburova, Del Popolo (2014) (squares). The left hand panel corresponds to piso dark halo, the right one -- to Burkert halo. The error bars for the sample of giant galaxies are associated with the range covered by $1\sigma$ confidence limit. The line shows the linear fit to the data.}\label{rc} 
\end{figure*}

\begin{figure*}
\includegraphics[width=1\textwidth]{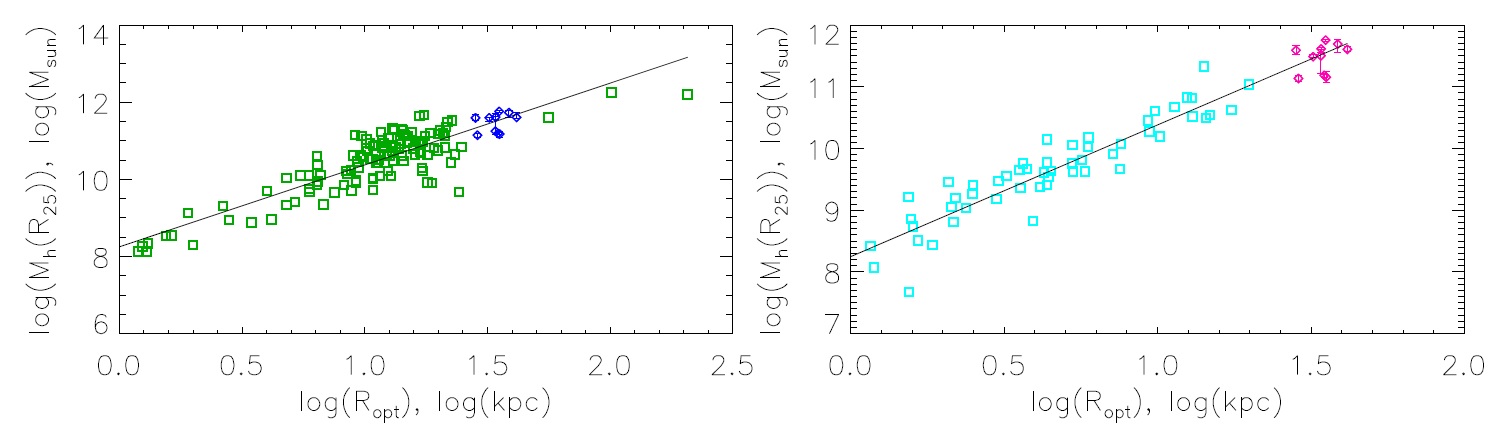}
\caption{ The comparison of the mass of the halo within disc radius with the optical radius for the current sample of giant spirals (diamonds) and the sample  from  Saburova, Del Popolo (2014) (squares). The left hand panel corresponds to piso dark halo, the right one -- to Burkert halo. The error bars for the sample of giant galaxies are associated with the range covered by $1\sigma$ confidence limit. The line shows the linear fit to the data.}\label{mhalo} 
\end{figure*}

\section{Conclusions}\label{conclusion} 
I selected a sample of giant discy galaxies which have optical radii higher than 30 kpc and are not tightly interacting. I compared the properties of these systems with that of the discy galaxies available in Hyperleda database. I also choose a subsample of 18 giant HSBs with available rotation curves and surface photometry data. For these objects, I performed the mass-modeliing of the rotation curves and constructed the $\chi^2$ maps for the parameters of dark matter haloes. The baryonic densities were fixed in a narrow ranges found from the photometry and stellar kinematical data (when they were available). I came to the following main conclusions:
\begin{itemize}
\item The rotation velocities, B-band luminosities, \HI masses and dynamical mass-to-light ratios of giant discy HSBs appear to be higher than that of the galaxies of moderate sizes. The latter can indicate the higher dark-to-luminous mass ratio. The dark matter could be either in non-baryonic or baryonic form (e.g. cold gas  non-detected by its emission).

\item The giant discs do not differ by the \HI mass to B-band luminosity ratio. It speaks  against the difference in star formation efficiency of giant spirals and moderate size galaxies.
\item I found that that the rings and bars occur more frequently in giant discy galaxies in comparison to moderate size spirals. It can imply that some of these systems are similar to giant ring galaxy UGC7069 studied by \cite{Ghosh2008}. 
\item Giant spirals follow the trend $\log(M_{\rm HI})(R_{25})$ found for normal size galaxies. This conclusion together with higher \HI masses and velocities of giant spirals can testify against the formation scenario of giant galaxies in which normal size galaxy is blown up. In this case one would expect giant galaxies to have similar masses as moderate size spirals and to lie below the correlation $\log(M_{\rm HI})(R_{25})$ which is disproved in current paper. 

It can also give evidences against  the peculiarities in the evolution of star formation of giant spirals, e.g. very strong star formation bursts or strangulation.

\item  I found that giant discy galaxies belong to clusters and groups slightly less frequently than normal-size galaxies. I propose that the most important role in the formation of giant discs should play not the current environment but that at the beginning of their evolution. 

\item Other specialties of giant discs are their shortage at the distances D < 100 Mpc and their lower mean B-band
surface brightnesses of the discs. The latter can indicate
that giant discs are more rarified in comparison to typical late type
galaxies. 

\item The dark halo mass and radial scale increases with the radius of the disc. These relations (at least the relation of sizes) are expected in the models of the formation of galaxies (se e.g. \citealt{Moetal1998}).  Giant HSBs have higher dark halo radial scales and masses than moderate size galaxies, but they do not deviate significantly from the trend of the normal size galaxies. It can indicate the absence of catastrophic scenario of the formation of giant galaxies. This conclusion also follows from the marginal gravitational stability of several giant discs with available stellar kinematical data.  All these findings support the formation scenario proposed by \citet{Kasparova2014} in which the  giant disc is formed due to the rarified dark halo with high radial scalelength.

\end{itemize}

\section*{Acknowledgements}  
I thank the anonymous referee for valuable comments that helped to improve the manuscript. 
I am grateful to Anatoly Zasov for fruitful discussion and his important comments on the manuscript. 
I thank my family for the support and inspiration. 

This work is supported by The Russian Science Foundation (RSCF) grant No. 17-72-20119.
I acknowledge the usage of the HyperLeda database (http://leda.univ-lyon1.fr).
\bibliographystyle{mnras}
\bibliography{Saburova}

\label{lastpage}

\end{document}